\documentclass[11pt,a4paper,oneside]{article}
\usepackage[utf8]{inputenc}
\usepackage[dvipsnames]{xcolor}
\usepackage[colorlinks=true,
            linkcolor=blue,
            urlcolor=green,
            citecolor=ForestGreen]{hyperref}
\usepackage{amsmath}
\usepackage{amsfonts}
\usepackage{amssymb}
\usepackage{indentfirst}
\usepackage{graphicx}
\usepackage{cleveref}
\DeclareMathOperator{\sign}{sign}
\graphicspath{{figures/}}
\usepackage{authblk}
\title{An Implicit High-Order Preconditioned Flux Reconstruction Method for Low-Mach-Number Flow Simulation with Dynamic Meshes}
\author{Lai Wang\thanks{Graduate student. Email: bx58858@umbc.edu} \ and Meilin Yu\thanks{Assistant Professor. Corresponding Author. Email: mlyu@umbc.edu}
}
\affil{Department of Mechanical Engineering, \\
University of Maryland, Baltimore County, Baltimore, MD 21250}
\date{\vspace{-10ex}}
\begin{document}
\maketitle
	\section*{Abstract}
	A fully implicit high-order preconditioned flux reconstruction/correction procedure via reconstruction (FR/CPR) method is developed to solve the compressible Navier--Stokes equations at low Mach numbers. A dual-time stepping approach with the second-order backward differentiation formula (BDF2) is employed to ensure temporal accuracy for unsteady flow simulation. When dynamic meshes are used to handle moving/deforming domains, the geometric conservation law (GCL) is implicitly enforced to eliminate errors due to the resolution discrepancy between BDF2 and the spatial FR/CPR discretization. The large linear system resulted from the spatial and temporal discretizations is tackled with the restarted Generalized Minimal Residual (GMRES) solver in the PETSc (Portable, Extensible Toolkit for Scientific Computation) library. Through several benchmark steady and unsteady numerical tests, the preconditioned FR/CPR methods have demonstrated good convergence and accuracy for simulating flows at low Mach numbers. The new flow solver is then used to study the effects of Mach number on unsteady force generation over a plunging airfoil when operating in low-Mach-number flows. It is observed that weak compressibility has a significant impact on thrust generation but a negligible effect on lift generation of an oscillating airfoil.

	\section*{Key Words}
	Flux Reconstruction/Correction Procedure via Reconstruction; Low-Mach-Number Preconditioning; High-Order Methods; Dual-Time Stepping; Unsteady Flows; Dynamic Meshes; Weak Compressibility
	\section{Introduction}
	Analysis of unsteady fluid flows at low Mach numbers is an indispensable component in many mechanical engineering applications, such as design and control of wind turbines, micro air vehicles, autonomous underwater vehicles, and medical devices, just to name a few.   
	There are typically two approaches to handle low-Mach-number flows. In the first one, fluids are considered incompressible, and their motion is governed by the incompressible Navier--Stokes equations. The pressure-correction method (e.g., SIMPLE (Semi-Implicit Method for Pressure Linked Equations) families~\cite{patankar1980numerical}, and the projection method~\cite{chorin1968numerical}), and artificial compressibility method~\cite{chorin1967numerical,Helenbrook06,bassi2006artificial,crivellini2013high,cox2016high,Yu2016} are among the most popular approaches for solving incompressible flows. 
	An alternative is to directly solve the compressible Navier--Stokes equations at low Mach numbers. 
	In this study, we pursue the second approach to solve low-Mach-number flows. We note that the compressible flow equations become stiff to solve numerically at low Mach numbers due to the large disparity between the flow (or entropy characteristic) velocity and the speed of sound. This disparity not only degrades the convergence of numerical solvers for the compressible Navier--Stokes equations, but also results in loss of numerical accuracy~\cite{TurkelLeer1993}.
	
    Various local preconditioning methods have been proposed to decrease the stiffness of the compressible Navier--Stokes equations at low Mach numbers. In general, after local preconditioning, the physical acoustic wave speeds are modified to be of the same order of magnitude as the flow speeds. Thus, the convergence rate of the flow solver can be greatly improved. Three major groups of preconditioning methods have been reported in the literatures~\cite{Colin2011}. Inspired by the artificial compressibility method developed by Chorin~\cite{chorin1967numerical}, Turkel adopted $ (p,\boldsymbol{v},dS) $, where $p$ is the pressure, $\boldsymbol{v}$ is the velocity vector, and $dS$ is the change of entropy, as working variables in a series of work~\cite{TurkelLeer1993, Turkel1987, Turkel1997, Turkel_1999_review}. Choi and Merkle ~\cite{ChoiMerkle1993}, and Weiss and Smith~\cite{ WeissSmith1995} employed $ (p,\boldsymbol{v},T) $ as working variables. Herein, $T$ is the temperature.  
    Van Leer \textit{et al.}~\cite{VanLeerRoe1991,LeeLeer1993} developed a symmetric Van Leer–Lee–Roe preconditioner using $ (dp/(\rho c),d\boldsymbol{v},dS) $ as working variables, where $\rho$ is the fluid density, and $c$ is the speed of sound. 
    We note that different sets of working variables have been assessed by Turkel \textit{et al.}~\cite{Turkel1997}. Hauke and Hughes in Ref.~\cite {Hauke1998} pointed out that the conservative incompressible formulation is well defined only for the entropy variables and the primitive variables. 
    As reported in Refs.~\cite{ChoiMerkle1993, WeissSmith1995,VenkaMerkle1995, Vigneron2006, Bassi2009}, direct approximations of the gradients of primitive variables can be more accurate than those computed from gradients of conservative variables. Therefore, in this study, the primitive variables are selected as working variables, and the preconditioning approach by Weiss and Smith~\cite{ WeissSmith1995} is adopted. Preconditioned high-order methods have been employed to solve the compressible Navier--Stokes equations at low Mach numbers in past decades. For example, the discontinuous Galerkin (DG) method has been developed to solve low-Mach-number flows in Refs.~\cite{Luo2008,Feistauer2007,Bassi2009,Nigro2010,Renda2015}. In addition to local preconditioning methods, we note that researchers have also proposed to directly modify the approximate Riemann solvers~\cite{thornber2008improved,rieper2011low,chen2018improved,chen2018effective} to balance numerical dissipation on different characteristic waves. A comprehensive comparison of several approximate Riemann solvers at low Mach numbers in the context of high-order flux reconstruction methods can be found in Ref.~\cite{boxi2017study}.

	\textit{Contributions}.  In this paper, an implicit preconditioned FR/CPR method is developed for the first time to solve the unsteady compressible Navier--Stokes equations at low Mach numbers involving dynamic meshes. 
	The dual-time stepping method with BDF2 is employed to achieve second-order accuracy in time for unsteady flow simulations. Within the framework of dual-time stepping, the GCL is enforced analytically following the approach proposed in Refs.~\cite{Yu2011,Liang14}. In a previous work~\cite{Wang_Yu2017}, we have compared the numerical performance of the preconditioned FR/CPR formulation with the incompressible FR/CPR method with artificial compressibility~\cite{Yu2016}. In this study, we conduct thorough numerical experiments to test accuracy, convergence and efficiency of the preconditioned FR/CPR formulation for solving steady and unsteady flow problems.
	
	\textit{Article overview}. The remainder of the paper is organized as follows. In~\Cref{sec:LMNPre}, the non-dimensionalized compressible Navier--Stokes equations are presented first (~\Cref{sec:governing_equations}). The preconditioning methodology for steady flow problems is then explained in~\Cref{sec:pre_steady}. After that, preconditioning for unsteady flow problems on moving grids with implicit GCL enforcement is elaborated in~\Cref{sec:pre_moving}.  Numerical methods are introduced in~\Cref{sec:NumMethods}. A brief review of the FR/CPR method is given in~\Cref{sec:spatial}, and~\Cref{sec:dual_time_stepping} formulates the concrete discrete forms of time marching methods. Boundary conditions for preconditioned systems are then discussed in~\Cref{sec:bc_conditions}. In~\Cref{sec:numerical_tests}, numerical results from both steady and unsteady simulations using the preconditioned FR/CPR methods are presented. The present study is summarized in~\Cref{sec:conclusions}.
	
	\section{Low-Mach-Number Preconditioning}\label{sec:LMNPre}
	
	\subsection{Governing Equations} \label{sec:governing_equations}
	The non-dimensionalized compressible Navier--Stokes equations can be written as 
	\begin{equation} \label{comp_ns}
	\frac{\partial \boldsymbol{q}^{c}}{\partial t} + \frac{\partial \boldsymbol{f}}{\partial x} +   \frac{\partial \boldsymbol{g}}{\partial y} = 0,
	\end{equation}
	with the following non-dimensionalizations~\cite{Pletcher1993},
	\begin{equation}
	\begin{cases}
	t={\bar{t}}/(L_{ref}/u_{ref}),\quad x={\bar{x}}/{L_{ref}},\\
	\rho={\bar{\rho}}/{\rho_{ref}}, \quad \boldsymbol{v}={\bar{\boldsymbol{v}}}/{u_{ref}}, \quad p={\bar{p}}/{\rho_{ref}u_{ref}^2},\quad T={\bar{T}}/{T_{ref}},\\
	\mu={\bar{\mu}}/{\mu_{ref}},\quad R =1/(\gamma Ma_{ref}^2), \quad C_p=1/((\gamma-1)Ma_{ref}^2).
	\end{cases}	
	\end{equation}
	Herein, $(\bar{*})$ is the variable in the physical domain, $(*_{ref})$ is the reference variable, $ \boldsymbol{q}^{c} = (\rho, \rho u, \rho v, E)^\top$ are the non-dimensionalized conservative variables, $\rho$ is the non-dimensionalized density, $\boldsymbol{v}=(u,v)^\top$ is the non-dimensionalized velocity vector, $ p=\rho RT $ is the non-dimensionalized pressure, $ T $ is the non-dimensionalized temperature, $ E=\frac{p}{\gamma-1}+\frac{1}{2}\rho(u^2+v^2) $ is the non-dimensionalized total energy, the heat capacity ratio $\gamma = 1.4 $, and $Ma$ is the Mach number. The fluxes $ \boldsymbol{f} $ and $ \boldsymbol{g} $ consist of inviscid and viscous parts, i.e., $ \boldsymbol{f} = \boldsymbol{f}_{inv}-\boldsymbol{f}_{vis} $ and $\boldsymbol{g} = \boldsymbol{g}_{inv}-\boldsymbol{g}_{vis} $. The components in $ \boldsymbol{f}_{inv}$, $\boldsymbol{g}_{inv}$, $\boldsymbol{f}_{vis}$, $\boldsymbol{g}_{vis} $ can be found below,
	\begin{equation}\label{convection}
	\boldsymbol{f}_{inv} =
	\begin{pmatrix}
	\rho u         \\
	\rho u^{2}+p   \\
	\rho uv        \\
	(E+p)u 
	\end{pmatrix},
	\boldsymbol{g}_{inv} =
	\begin{pmatrix}
	\rho v         \\
	\rho uv   \\
	\rho v^{2}+p        \\
	(E+p)v 
	\end{pmatrix},		 
	\end{equation}
	\begin{equation}\label{diffusion}
	\boldsymbol{f}_{v} =\frac{1}{Re_{ref}}
	\begin{pmatrix}
	0         \\
	\tau_{xx}   \\
	\tau_{xy}        \\
	u\tau_{xx}+v\tau_{xy}+k\frac{\partial T}{\partial x} 
	\end{pmatrix},
	\boldsymbol{g}_{v} =\frac{1}{Re_{ref}}
	\begin{pmatrix}
	0         \\
	\tau_{xy}   \\
	\tau_{yy}        \\
	u\tau_{xy}+v\tau_{yy}+k\frac{\partial T}{\partial y}
	\end{pmatrix}.		 
	\end{equation}	
	Herein, the non-dimensionalized thermal conductivity $ k = \frac{\mu C_{p} }{Pr}$, the Prandtl number $ Pr=0.72 $ and $ \mu $ is the non-dimensionalized dynamic viscosity, which is treated as a constant in this work. $ Re_{ref}=\rho_{ref}u_{ref}L_{ref}/\mu_{ref} $ is the Reynolds number of the reference flow at far fields. Non-dimensionalized viscous stresses are given by
	\begin{equation}\label{shear_stress}
	\begin{cases}
	\tau_{xx} = \mu(\frac{4}{3}\frac{\partial u}{\partial x}-\frac{2}{3}\frac{\partial v}{\partial y}),  \\
	\tau_{yy} = \mu(\frac{4}{3}\frac{\partial v}{\partial y}-\frac{2}{3}\frac{\partial u}{\partial x}),  \\
	\tau_{xy} = \tau_{yx} = \mu(\frac{\partial u}{\partial y}+\frac{\partial v}{\partial x}). 
	\end{cases}
	\end{equation}
	
	At low-Mach-number regions, the eigenvalues of the convection part of Eq.~\eqref{comp_ns} 
	are significantly different from each other. This disparity makes the hyperbolic system stiff to solve. Preconditioning methods replace the characteristic wave speeds with modified ones to decrease the stiffness of the hyperbolic system.
	
	\subsection{Preconditioning for Steady Flow Problems}\label{sec:pre_steady}
	A pseudo transient continuation approach is used to handle steady flow problems. In this approach, on applying the chain-rule, Eq.~\eqref{comp_ns} can be written as  
	\begin{equation} \label{changed_ns}
	\frac{\partial \boldsymbol{q}^{c}}{\partial \boldsymbol{q}^{p}}\frac{\partial \boldsymbol{q}^{p}}{\partial \tau} + \frac{\partial \boldsymbol{f}}{\partial x} +   \frac{\partial \boldsymbol{g}}{\partial y} = 0,
	\end{equation}
	where 
	$\tau$ is the pseudo-time, 
	$\boldsymbol{q}^{p} = (p, u, v, T)^\top$ are the non-dimensionalized primitive variables, $\boldsymbol{ M} = \frac{\partial \boldsymbol{q}^{c}}{\partial \boldsymbol{q}^{p}} $ is the Jacobian matrix of the two sets of variables, i.e., the conservative and primitive variables, which can be expressed as
	\begin{equation}\label{jacobian}
	\boldsymbol{M} = \frac{\partial \boldsymbol{q}^{c}}{\partial \boldsymbol{q}^{p}} =
				\begin{pmatrix}
					\rho_{p}    & 0      & 0      & \rho_{T}   \\
					\rho_{p}u   & \rho   & 0      & \rho_{T} u \\
					\rho_{p}v   & 0      & \rho   & \rho_{T} v \\
					\rho_{p}h-1 & \rho u & \rho v & \rho_{T} h + \rho C_{p} 
				 \end{pmatrix}.			 
	\end{equation}
	Herein, $h$ is the non-dimensionalized specific total enthalpy defined as $ h=C_{p}T+\frac{1}{2}(u^{2}+v^{2}) $.
	
	Recall that the state equation $ p=\rho RT $. 
	Thus the derivatives in Eq.~\eqref{jacobian} can be calculated as~\cite{Bassi2009} 
	\begin{equation}
	\rho_{p} = \frac{\partial \rho}{\partial p}\Bigr|_{T} = \frac{1}{RT},
	\rho_{T} = \frac{\partial \rho}{\partial T}\Bigr|_{p} = -\frac{\rho}{T}.
	\end{equation}	
	
	In the preconditioning method adopted here, $ \boldsymbol{M} $ is replaced with a matrix $ \boldsymbol{\varGamma} $ called preconditioning matrix, which can cluster the eigenvalues of the hyperbolic system. The preconditioning matrix is given by
	\begin{equation}\label{preconditioning}
	\boldsymbol{\varGamma} =
	\begin{pmatrix}
	\Theta    & 0      & 0      & \rho_{T}   \\
	\Theta u  & \rho   & 0      & \rho_{T} u \\
	\Theta v  & 0      & \rho   & \rho_{T} v \\
	\Theta h-1& \rho u & \rho v & \rho_{T} h + \rho C_{p} 
	\end{pmatrix}.			 
	\end{equation} 
	Note that $ \rho_{p} = \frac{\partial \rho}{\partial p} $  in $ \boldsymbol{M}$ is substituted with   $ \varTheta  $ in the preconditioning matrix $ \boldsymbol{\varGamma} $. The definition of $ \varTheta  $ follows that in Ref.~\cite{WeissSmith1995}
	\begin{equation}\label{theta}
	\varTheta = \Bigl(\frac{1}{U^{2}_{r}}-\frac{\rho_{T}}{\rho C_{p}}\Bigr),
	\end{equation}
	where $ U_{r} = \epsilon c$, $c$ is the local speed of sound, and  $ \epsilon $ is a free parameter related to the local Mach number $ Ma $, and the global cut-off Mach number which is usually chosen as the free stream Mach number $ Ma_\infty $.
	
	After introducing the preconditioning matrix into the compressible Navier--Stokes equations, the governing equations can be re-expressed as 
	\begin{equation} \label{steady_pre}
	\boldsymbol{\varGamma} \frac{\partial \boldsymbol{q}^{p}}{\partial \tau} + \frac{\partial \boldsymbol{f}}{\partial x} +   \frac{\partial \boldsymbol{g}}{\partial y} = 0.
	\end{equation}
	The preconditioned Jacobian matrix related to inviscid fluxes in the normal direction of any surface becomes $ \boldsymbol{\varGamma }^{-1}\boldsymbol{A}_{\boldsymbol{n}}$,  where $\boldsymbol{A}_{\boldsymbol{n}}=n_{x}\frac{\partial \boldsymbol{f}_{inv}}{\partial \boldsymbol{q}^{p}}+n_{y}\frac{\partial \boldsymbol{g}_{inv}}{\partial \boldsymbol{q}^{p}}$, and $\boldsymbol{n}=(n_{x},n_{y})^\top$ is the unit normal vector of the surface. The eigenvalues of the preconditioned Jacobian matrix in the normal direction are~\cite{WeissSmith1995}
	\begin{equation}\label{eigenvalues}
	\begin{cases}
		\lambda_{1} = \lambda_{2} = u_n \\
		\lambda_{3} = u'_n+c'\\
		\lambda_{4} = u'_n-c',
	\end{cases}
	\end{equation}
	where
	\begin{equation}\label{eigenvalues_para}
	\begin{cases}
	u_n=\boldsymbol{v}\cdot\boldsymbol{n}\\
	u'_n=u_n(1-\alpha)\\
	c'=\sqrt{\alpha^{2}u^{2}_n+U_{r}^{2}} \\
	\alpha = (1-\beta U_{r}^{2})/2\\
	\beta = \left(\rho_{p}+\frac{\rho_{T}}{\rho C_{p}}\right).
	\end{cases}
	\end{equation}
	For an ideal gas, $ \beta=(\gamma RT)^{-1}=1/c^{2} $. At low speed, when $ U_{r} $ approaches zero, $ \alpha $ will approach $ \frac{1}{2} $. All the eigenvalues will then have the same magnitude as $ u_n $. Thus, the stiffness of the compressible Navier--Stokes equations is significantly decreased.
	
	In the present study, when no moving grids are involved, the parameter $ \epsilon $ in $ U_r=\epsilon c $ is defined as
	\begin{equation}\label{eps_nonmove}
	\epsilon = \min(1,\max(\kappa Ma_\infty, Ma))
	\end{equation} 
	where $ \kappa$ is a free parameter. Typically, $ \kappa = 1 $. Turkel suggested that $ \kappa $ can be up to $ \sqrt{3}\sim \sqrt{5} $ in Ref.~\cite{Turkel_1999_review}. The global cut-off parameter  $\kappa Ma_\infty $~\cite{VenkaMerkle1995,Turkel_1999_review} is employed to prevent robustness deterioration instabilities near stagnation points.

	\subsection{Preconditioning for Unsteady Flow Problems with Dynamic Meshes}\label{sec:pre_moving}
	The preconditioning matrix $ \boldsymbol{\varGamma} $ in Eq.~\eqref{changed_ns} destroys the temporal accuracy of Eq.~\eqref{comp_ns}. The dual-time stepping method~\cite{Jameson1991} formulated as 
	\begin{equation} \label{unsteady_pre_nomov}
	\boldsymbol{\varGamma} \frac{\partial \boldsymbol{q}^{p}}{\partial \tau}+\frac{\partial \boldsymbol{q}^{c}}{\partial t} + \frac{\partial \boldsymbol{f}}{\partial x} +   \frac{\partial \boldsymbol{g}}{\partial y}= 0,
	\end{equation}
	can be an intuitive solution for preconditioned unsteady flow problems.
	In this work, the BDF2 method is employed to discretize the physical time derivative term, thus achieving a second-order accuracy temporally. When dynamic meshes are involved, in order to facilitate the incorporation of GCL, one can transfer Eq.~\eqref{unsteady_pre_nomov} from the physical domain $ (t,x,y) $ into the computational domain $ (t^{*}, \xi, \eta) $ as
	\begin{equation} \label{unsteady_pre_mov_cdomain}
	\boldsymbol{\varGamma} |\boldsymbol{J}|\frac{\partial \boldsymbol{q}^{p}}{\partial \tau}+\frac{\partial |
	\boldsymbol{J}|\boldsymbol{q}^{c}}{\partial t^*} + \frac{\partial \boldsymbol{F}}{\partial \xi} +   \frac{\partial \boldsymbol{G}}{\partial \eta}= 0,
	\end{equation}	
	where
	\begin{equation}
	\begin{cases}\label{computational_flux}
	\boldsymbol{F} =\left|\boldsymbol{J}\right| \left(\boldsymbol{q}^{c}\xi_t+\boldsymbol{f}\xi_{x}+\boldsymbol{g}\xi_{y}\right) \\
	\boldsymbol{G}=\left|\boldsymbol{J}\right| \left(\boldsymbol{q}^{c}\eta_t+\boldsymbol{f}\eta_{x}+\boldsymbol{g}\eta_{y}\right).
	\end{cases}
	\end{equation}
	The GCL of the transformation can be formulated as 
	\begin{equation}\label{gcl_eqs}
	\begin{cases}
	\frac{\partial}{\partial \xi}(|\boldsymbol{J}|\xi_x)+\frac{\partial}{\partial \eta}(|\boldsymbol{J}|\eta_x)=0 \\
	\frac{\partial}{\partial \xi}(|\boldsymbol{J}|\xi_y)+\frac{\partial}{\partial \eta}(|\boldsymbol{J}|\eta_y)=0 \\
	\frac{\partial |\boldsymbol{J}|}{\partial t^*}+\frac{\partial}{\partial \xi}(|\boldsymbol{J}|\xi_t)+\frac{\partial}{\partial \eta}(|\boldsymbol{J}|\eta_t)=0.
	\end{cases}
	\end{equation}
	In this study, $ t^* = t $, thus,
	\begin{equation}
	\boldsymbol{J} = \frac{\partial(x,y,t)}{\partial(\xi,\eta,t^*)}=
	\begin{pmatrix}
	x_\xi & x_\eta & x_{t^*}\\
	y_\xi & y_\eta & y_{t^*}\\
	0 & 0 & 1\\
	\end{pmatrix}
	\end{equation}
	and 
	\begin{equation}
	\quad|\boldsymbol{J}| = x_{\xi}y_{\eta}-x_{\eta}y_{\xi}, \quad
	\begin{cases}
	\xi_t = -\boldsymbol{v}_g\cdot\nabla\xi,\\
	\eta_t = -\boldsymbol{v}_g\cdot\nabla\eta.\\
	\end{cases}
	\end{equation}
	Herein, $ \boldsymbol{v}_g=(u_{g},v_g)^\top $ is the grid velocity vector. Since grid velocities are only related to the physical time $ t $ or $ t^* $, when marching along the pseudo-time $ \tau $, GCL can be enforced analytically to eliminate the error discrepancy between BDF2 and the spatial FR/CPR schemes following the approach in Refs.~\cite{Yu2016,Yu2011,Liang14}. Specifically, on substituting the last equality in Eq.~\eqref{gcl_eqs} into Eq.~\eqref{unsteady_pre_mov_cdomain}, Eq.~\eqref{unsteady_pre_mov_cdomain} can be written as
	\begin{equation} \label{unsteady_pre_mov_cdomain_GCL}
		\boldsymbol{\varGamma} |\boldsymbol{J}|\frac{\partial \boldsymbol{q}^{p}}{\partial \tau} + 
		|\boldsymbol{J}| \frac{\partial \boldsymbol{q}^{c}}{\partial t^*} -
		\boldsymbol{q}^{c} \left( \frac{\partial |\boldsymbol{J}| \xi_t}{\partial \xi} +
		\frac{\partial |\boldsymbol{J}| \eta_t}{\partial \eta}\right)
		+ \frac{\partial \boldsymbol{F}}{\partial \xi} +   \frac{\partial \boldsymbol{G}}{\partial \eta}= 0,
		\end{equation}
	One can further substitute Eq.~\eqref{computational_flux} into the above equation and transfer it back  onto the physical domain using the first two formulae in Eq.~\eqref{gcl_eqs}, which  reads 
	\begin{equation} \label{unsteady_pre_mov_pdomain}
	\boldsymbol{\varGamma} \frac{\partial \boldsymbol{q}^{p}}{\partial \tau}+\frac{\partial\boldsymbol{q}^{c}}{\partial t^*} + \frac{\partial \boldsymbol{f}}{\partial x} +   \frac{\partial \boldsymbol{g}}{\partial y}-u_g\frac{\partial \boldsymbol{q}^c}{\partial x}-v_g\frac{\partial \boldsymbol{q}^c}{\partial y}= 0.
	\end{equation}
	Therefore, the GCL is analytically enforced. The error are only related to the numerical discretizations.
	Interested readers are referred to our previous work~\cite{Yu2016} for a more detailed derivation. There the governing equations for incompressible flows on dynamic meshes are solved with FR/CPR via the artificial compressibility approach.

    When dynamic meshes are involved, the preconditioned Jacobian matrix in the normal direction of any surface for the convection terms becomes $ \boldsymbol{\varGamma}^{-1}\boldsymbol{A}_{\boldsymbol{n},mov} $, where
    \begin{equation}
    \boldsymbol{A}_{\boldsymbol{n},mov} = n_{x}\frac{\partial \boldsymbol{f}_{inv}}{\partial \boldsymbol{q}^{p}}+n_{y}\frac{\partial \boldsymbol{g}_{inv}}{\partial \boldsymbol{q}^{p}}-u_{g}\frac{\partial \boldsymbol{q}^c}{\partial \boldsymbol{q}^{p}}n_x -v_{g}\frac{\partial \boldsymbol{q}^c}{\partial \boldsymbol{q}^{p}}n_y.
    \end{equation}
    The eigenvalues of this system preserve the same form as those in Eq.~\eqref{eigenvalues} or Eq.~\eqref{eigenvalues_para} except a minor change, i.e., $ u_n $ in Eq.~\eqref{eigenvalues_para} is modified to
	\begin{equation}
	u_n=(\boldsymbol{v}-\boldsymbol{v}_g)\cdot\boldsymbol{n},
	\end{equation}
	due to the presence of the grid velocity.
	The preconditioning parameter $ \epsilon $ for dynamic meshes is still defined as~\cite{Xiao2007}
	\begin{equation}\label{eps_move}
	\epsilon = \min(1,\max(\kappa Ma_\infty, Ma_{mov})),
	\end{equation}
	where $ Ma_{mov} $ is the Mach number calculated from $ \boldsymbol{v}-\boldsymbol{v}_g $.
	
	\section{Numerical Methods}\label{sec:NumMethods}
	
	\subsection{Spatial Discretization Method} \label{sec:spatial}
	
	For completeness, a brief review of the FR/CPR method~\cite{Huynh2007, Huynh2009, ZJWang2009, Vincent2011} is presented in this section. In FR/CPR, fluxes can be divided into two parts, namely, local fluxes constructed directly from local solutions, and correction fluxes accounting for the differences between the local fluxes and common fluxes constructed from those on element interfaces. In the current study, the spatial domain is discretized into non-overlapping quadrilateral elements.  When the FR/CPR method is employed for  the spatial discretization,  
	Eq.~\eqref{unsteady_pre_mov_pdomain} can be reformulated as 
	\begin{equation} \label{cpr_ns}
	\begin{aligned}
	\boldsymbol{\varGamma} \frac{\partial \boldsymbol{q}^{p}}{\partial \tau} &+ \frac{\partial \boldsymbol{f}^{loc}}{\partial x} +   \frac{\partial \boldsymbol{g}^{loc}}{\partial y}-u_{g}\frac{\partial \boldsymbol{q}^{c,loc}}{\partial x} -v_{g}\frac{\partial \boldsymbol{q}^{c,loc}}{\partial y}\\ &+\frac{1}{\left|\boldsymbol{J}\right|}\left(\frac{\partial \boldsymbol{F}^{cor}}{\partial \xi} +   \frac{\partial \boldsymbol{G}^{cor}}{\partial \eta}\right) + \frac{\partial \boldsymbol{q}^c}{\partial t^*} = 0.
	\end{aligned}	
	\end{equation} 
	For quadrilateral elements, $ F^{cor} $ and $ G^{cor} $ are given by
	\begin{equation}
	\begin{cases}
	\begin{aligned}
	\boldsymbol{F}^{cor}(\xi, \eta) &= \left(\boldsymbol{F}_{L}^{num}(\eta)-\boldsymbol{F}^{loc}(\xi_{L}, \eta)\right)g_{L}^{c}(\xi)\\
	&+\left(\boldsymbol{F}_{R}^{num}(\eta)-\boldsymbol{F}^{loc}(\xi_{R}, \eta)\right)g_{R}^{c}(\xi),
	\end{aligned}\\
	\begin{aligned}
	\boldsymbol{G}^{cor}(\xi, \eta) &= \left(\boldsymbol{G}_{L}^{num}(\xi)-\boldsymbol{G}^{loc}(\xi, \eta_{L})\right)g_{L}^{c}(\eta)\\
	&+\left(\boldsymbol{G}_{R}^{num}(\xi)-\boldsymbol{G}^{loc}(\xi, \eta_{R})\right)g_{R}^{c}(\eta),
	\end{aligned}
	\end{cases}
	\end{equation}
	where $ \boldsymbol{F}^{num} $ and $ \boldsymbol{G}^{num} $  are the common (or numerical) fluxes on the interfaces of standard elements.
	Herein, subscripts `\textit{L}' and `\textit{R}' stand for left and right edges of an element in a dimension by dimension sense. $ g_{L}^{c}, g_{R}^{c} $ are correction functions which map the differences between the common and local fluxes on the element interfaces into the entire standard element. In this study, $ g_{L}^{c}$ and $g_{R}^{c} $ are the right and left Radau polynomials, which can recover the nodal DG methods.  Note that common fluxes are always calculated in the physical domain. Thus $ \boldsymbol{F}^{num} $ and $ \boldsymbol{G}^{num} $ can be expresses as
	\begin{equation}\label{norm_f_2}
	\begin{cases}
	\boldsymbol{F}^{num}={|\boldsymbol{J}|}|\nabla\xi|\boldsymbol{f}^{com}_{\boldsymbol{n}} \sign(\boldsymbol{n}\cdot\nabla\xi),\\
	\boldsymbol{G}^{num}={|\boldsymbol{J}|}|\nabla\eta|\boldsymbol{f}^{com}_{\boldsymbol{n}} \sign(\boldsymbol{n}\cdot\nabla\eta),
	\end{cases}
	\end{equation}
	where $ \boldsymbol{f}^{com}_{\boldsymbol{n}} $ are the common fluxes in  the normal  direction $ \boldsymbol{n} $ of a physical surface. On element interfaces, the inviscid fluxes along $ \boldsymbol{n} $ can be written as
	\begin{equation}
	\boldsymbol{f}_{\boldsymbol{n},inv} = \boldsymbol{f}n_x+\boldsymbol{g}n_y-\left(\boldsymbol{v}_g\cdot\boldsymbol{n}\right)\boldsymbol{q}^c.
	\end{equation}
	Therefore, $ \boldsymbol{f}^{com}_{\boldsymbol{n},inv} $  can be formulated as, taking the Rusanov approximate Riemann solver as an example,
	\begin{equation}\label{Rosanov}
	\boldsymbol{f}_{\boldsymbol{n},inv}^{com}=\frac{1}{2}\left(\boldsymbol{f}_{\boldsymbol{n},inv}^{+}+\boldsymbol{f}_{\boldsymbol{n},inv}^{-}\right)-\frac{1}{2}\left|\lambda\right|_{max}\boldsymbol{\widetilde{\varGamma}} \left(\boldsymbol{q}^{p,+}-\boldsymbol{q}^{p,-}\right).
	\end{equation}
	where $\boldsymbol{\widetilde{\varGamma}}$ is the local preconditioning matrix calculated by averaging primitive variables on elements interfaces, and superscripts `$ + $' and `$ - $' denote the right and left side of the interface, respectively. $\lambda$ stands for the eigenvalue of the Jacobian matrix $ \boldsymbol{\varGamma}^{-1}\boldsymbol{A}_{\boldsymbol{n}} $ or $ \boldsymbol{\varGamma}^{-1}\boldsymbol{A}_{\boldsymbol{n},mov} $.
	
	The preconditioning does not affect the viscous terms and $ \boldsymbol{f}_{n,vis}^{com} = \boldsymbol{f}_{vis}(\boldsymbol{q}^{p,+},\nabla \boldsymbol{q}^{p,+},\boldsymbol{q}^{p,-},\nabla \boldsymbol{q}^{p,-}) $. To calculate the common viscous fluxes, we need to define common $ \boldsymbol{q}^{p,com} $ and common $ \nabla \boldsymbol{q}^{p,com} $ on element interfaces. 
	Simply taking average of the primitive variables yields
	\begin{equation}
	\boldsymbol{q}^{p,com} = \frac{\boldsymbol{q}^{p,+}+\boldsymbol{q}^{p,-}}{2}.
	\end{equation}
	The common gradient is calculated following the the second approach of Bassi and Rebay (BR2) \cite{BR2_2005} as 
	\begin{equation}
	\nabla \boldsymbol{q}^{p,com} = \frac{\nabla\boldsymbol{q}^{p,+}+\boldsymbol{r}^{+}+\nabla\boldsymbol{q}^{p,-}+\boldsymbol{r}^{-}}{2},
	\end{equation}
	where $ \boldsymbol{r}^{+} $ and $ \boldsymbol{r}^{-} $ are the corrections to the gradients on the interfaces.
	See Ref.~\cite{Gao2013} for more information.
	
	\subsection{Time Marching Method}\label{sec:dual_time_stepping}
	For steady flow problems, define $ \boldsymbol{R}=-\left(\frac{\partial \boldsymbol{f}}{\partial x}+\frac{\partial \boldsymbol{g}}{\partial y}\right) $ and substitute it into Eq.~\eqref{steady_pre} to obtain
	\begin{equation} \label{steady_pre_2}
	\boldsymbol{\varGamma} \frac{\partial \boldsymbol{q}^{p}}{\partial \tau} = \boldsymbol{R}.
	\end{equation}
	A backward Euler method is used to discretize the pseudo-time derivative, and the residual $\boldsymbol{R}$ is linearized around $ \boldsymbol{R}^{m}$ at the pseudo-time step \textit{m}. As a result, the linear system to be solved becomes 
	\begin{equation} \label{steady_pre_discretized_1}	
	\left(\frac{\boldsymbol{\varGamma}^{m} }{\Delta \tau}-\left(\frac{\partial \boldsymbol{R}}{\partial \boldsymbol{q}^{p}}\right)^{m}\right)\Delta(\boldsymbol{q}^p)^{m} 
	= \boldsymbol{R}^{m},		
	\end{equation} 
	where $ \Delta\left(\boldsymbol{q}^p\right)^m = \left(\boldsymbol{q}^p\right)^{m+1}-\left(\boldsymbol{q}^p\right)^m $. A widely used approach to update $ \Delta \tau $ is the switched evolution relaxation (SER) originally developed by Mulder and van Leer~\cite{Mulder1985}, which reads
	\begin{equation}\label{SER}
	\Delta \tau_{m+1} = \max\left(\Delta \tau_{min}, \min\left(\Delta \tau_{max},\Delta \tau_{m} \left(\frac{Res_{m-1}/Res_0}{Res_m/Res_0}\right)^{r_{ser}}\right)\right),
	\end{equation}
	where $ Res_{m}/Res_{0} $ is the relative residual of pressure at the pseudo-time step $m$, and $ r_{ser} $ is a free parameter to control the adaptation rate of $ \Delta \tau $. If not specifically mentioned,  $\Delta \tau_{min}= \Delta \tau_{0} = 0.01$, $\Delta  \tau_{max}=10^{20} $, and $ r_{ser} = 2 $ are employed for all steady problems. We note that with $ r_{ser} = 2 $ the SER employed in this study is a relatively aggressive approach to update $ \Delta \tau $, which could lead to divergence of the  pseudo transient continuation. As will be demonstrated in Sections \ref{subsec:cylinder} and \ref{Steady_NACA12}, two approaches are recommended to facilitate convergence: in the first one, a tight (or small) tolerance of the linear solver, i.e. restarted GMRES, is used to ensure its convergence at each pseudo-time step; in the other, the maximum pseudo-time step size $\Delta  \tau_{max} $ is reduced. We note that Ceze and Fidkowski have proposed more sophisticated strategies for updating $ \Delta  \tau $ in Refs.~\cite{ceze2011robust,ceze2015constrained}. They found that  a constraint to avoid non-physical state is important to improve the robustness of the pseudo transient continuation. Interested readers are referred to their work; further discussion on this topic is beyond the scope of this paper.
	
	For unsteady flow problems with the dual-time stepping method,
	define $ \boldsymbol{R}=-\left(\frac{\partial \boldsymbol{f}}{\partial x}+\frac{\partial \boldsymbol{g}}{\partial y}-u_{g}\frac{\partial \boldsymbol{q}^c}{\partial x} -v_{g}\frac{\partial \boldsymbol{q}^c}{\partial y}\right) $. On plugging it into Eq.~\eqref{unsteady_pre_mov_pdomain}, we obtain 
	\begin{equation} \label{unsteady_pre_2}
	\boldsymbol{\varGamma} \frac{\partial \boldsymbol{q}^{p}}{\partial \tau} = -\frac{\partial \boldsymbol{q}^{c}}{\partial t^*}+\boldsymbol{R}.
	\end{equation}
	If $ n $ denotes the current physical-time step, and $m$ denotes the pseudo-time step, then we have
	\begin{equation} \label{unsteady_pre_discretized_1}
	\boldsymbol{\varGamma}^{m} \frac{(\boldsymbol{q}^p)^{m+1}-(\boldsymbol{q}^p)^{m}}{\Delta \tau} = -\left(\frac{\partial \boldsymbol{q}^{c}}{\partial t^*}\right)^{m+1}+\boldsymbol{R}^{m+1},
	\end{equation}
	where $ (\boldsymbol{q}^{p})^{m=0} = (\boldsymbol{q}^{p})^{n} $ and when $ m \to \infty $, $ (\boldsymbol{q}^{p})^{m+1} \to (\boldsymbol{q}^{p})^{n+1} $. To achieve a second-order temporal accuracy, BDF2 is used to discretize the derivative with respect to $ t^* $. Thus, Eq.~\eqref{unsteady_pre_discretized_1} is discretized as
	\begin{equation} \label{unsteady_pre_discretized_2}
	\boldsymbol{\varGamma}^{m} \frac{(\boldsymbol{q}^p)^{m+1}-(\boldsymbol{q}^p)^{m}}{\Delta \tau} = -\frac{3(\boldsymbol{q}^{c})^{m+1}-4(\boldsymbol{q}^{c})^{n}+(\boldsymbol{q}^{c})^{n-1}}{2 \Delta t}+\boldsymbol{R}^{m+1}.
	\end{equation}
	Note that $\Delta t = \Delta t^*$. After linearizing $\boldsymbol{R}^{m+1}$ with respect to $\boldsymbol{R}^{m}$, Eq.~\eqref{unsteady_pre_discretized_2} can be further written as 
	\begin{equation} \label{unsteady_pre_discretized_3}
	\begin{aligned}
	&\left(\frac{\boldsymbol{\varGamma}^{m} }{\Delta \tau}+\frac{3}{2 \Delta t}\boldsymbol{M}^{m}-\left(\frac{\partial \boldsymbol{R}}{\partial \boldsymbol{q}^{p}}\right)^{m}\right)\Delta(\boldsymbol{q}^p)^{m} \\
	&= -\frac{3(\boldsymbol{q}^{c})^{m}-4(\boldsymbol{q}^{c})^{n}+(\boldsymbol{q}^{c})^{n-1}}{2 \Delta t}+\boldsymbol{R}^{m},
	\end{aligned}	
	\end{equation}
	where $\Delta(\boldsymbol{q}^p)^{m}=(\boldsymbol{q}^p)^{m+1}-(\boldsymbol{q}^p)^{m}  $. For unsteady flow problems tested here, $ \Delta \tau$ is always fixed as $\Delta \tau = 10^{20} $ for pseudo-time iterations. 
	
	In the present study, the full matrix $ \frac{\partial \boldsymbol{R}}{\partial \boldsymbol{q}^{p}} $ is computed numerically using the first-order finite difference approach. Note that 
	if only block diagonal terms of the Jacobian matrix were supplied, the convergence would be dramatically deteriorated~\cite{Bassi2009}. The restarted GMRES solver in the PETSc library~\cite{petsc-user-ref} is used to solve Eq.~\eqref{steady_pre_discretized_1} and Eq.~\eqref{unsteady_pre_discretized_3}. 
	\subsection{Boundary Conditions}\label{sec:bc_conditions}
	The characteristics of the inviscid parts of the compressible Navier--Stokes equation are changed due to the low-Mach-number preconditioning. No-reflection far-field boundary conditions or simplified far-field boundary conditions have been proposed in Refs.~\cite{Turkel1997, Bassi2009}. However, the far field in this study is always located more than 100 characteristic lengths (e.g., the diameter of a cylinder, the chord length of an airfoil) away from the solid wall. Therefore, we simply use  $ \boldsymbol{q}^{p,b} = \boldsymbol{q}^{p,\infty} $ on boundaries at the far field, and enforce boundary conditions using the approximate Riemann solver.
	For inviscid flows, the slip wall boundary condition is employed.
	The adiabatic wall boundary condition is enforced for no-slip viscous walls. To ensure high-order accuracy on wall boundaries, $ P^4 $ mesh elements are employed to handle the curvatures of wall boundaries.
	\section{Numerical Results}\label{sec:numerical_tests}
	Several benchmark tests have been conducted to verify the accuracy and convergence of the preconditioned FR/CPR method in this section.  The $ L_2 $ error is used for the accuracy study. Specifically, for any solution variables $ \phi $ on domain $ \Omega $, the $ L_2 $ error is defined as
	 \begin{equation}
	 E_\phi = \sqrt{\frac{\int_{\Omega} \left(\phi-\phi^{exact}\right)^2dV}{\int_{\Omega}1dV}}.
	 \end{equation}
	 Exclusively, the entropy error $ E_s $ is defined as 
	 \begin{equation}
	  E_s = \sqrt{\frac{\int_{\Omega} \left(p/\rho^\gamma-p_{\infty}/\rho_\infty^\gamma\right)^2dV}{\int_{\Omega}(p_{\infty}/\rho_{\infty}^\gamma)^2 dV}}.
	 \end{equation}

	 To monitor the convergence of nonlinear systems, such as Eqs.~\eqref{steady_pre_2} and~\eqref{unsteady_pre_2}, two convergence criteria are set up for the pseudo-time iteration: one is for $ Res_{m}/Res_{0} $, denoted as $tol_{pseudo}$; the other is for the restarted GMRES linear solver, denoted as $tol_{res}$. For steady flow simulation using a pseudo transient continuation (see Eq.~\eqref{steady_pre_2}), the goal is to decrease the steady relative residual $Res_{m}/Res_{0}$ as much as possible with a reasonable criterion for $tol_{res}$; for unsteady flow simulation (see Eq.~\eqref{unsteady_pre_2}), the goal is to decrease the unsteady relative residual $Res_{m}/Res_{0}$ to a satisfactory level with a reasonable criterion for $tol_{res}$.     
	 
	 The drag coefficient $ C_d $, lift coefficient $ C_l $ and thrust coefficient $ C_t $ are defined as
	 \begin{equation}
	 C_d = \frac{Drag}{\frac{1}{2}\rho_{\infty}U_{\infty}^2 A}, \quad C_l = \frac{Lift}{\frac{1}{2}\rho_{\infty}U_{\infty}^2 A},\quad \text{and} \quad C_t = -C_d,
	 \end{equation}
	 where $A$ is the area that the force acts on, and $U_{\infty}$ is the free stream velocity.
	 The pressure is normalized as 
	 \begin{equation}
	 p_{norm} = \frac{p-p_{min}}{p_{max}-p_{min}},
	 \end{equation}
	 for better visualization, where $ p_{min} $ and $ p_{max} $ are the minimum and maximum pressure in the entire flow field.
	 
	 A parallel code is developed to accelerate the simulations. The block Jacobi preconditioner is employed for parallelization, and ILU(0) is used as the local preconditioner for the restarted GMRES solver in PETSc. Each process possesses only one mesh block~\cite{Bassi2015}. The dimension of the Krylov subspace is 150 for steady flow problems and 30 for unsteady ones. The Jacobian matrix $ \frac{\partial \boldsymbol{R}}{\partial \boldsymbol{q}^p} $ is updated every pseudo-time iteration for steady flow problems and every two pseudo-time iterations for unsteady flow problems. All the steady flow problems are simulated with one process, and all the unsteady cases are solved with 16 processes.

	\subsection{Inviscid Flow over a Circular Cylinder} \label{subsec:cylinder}
	\label{Steady_Cylinder}
\begin{figure}[t]
	\centering
	\begin{tabular}{cc}
	\includegraphics[width=6.5cm]{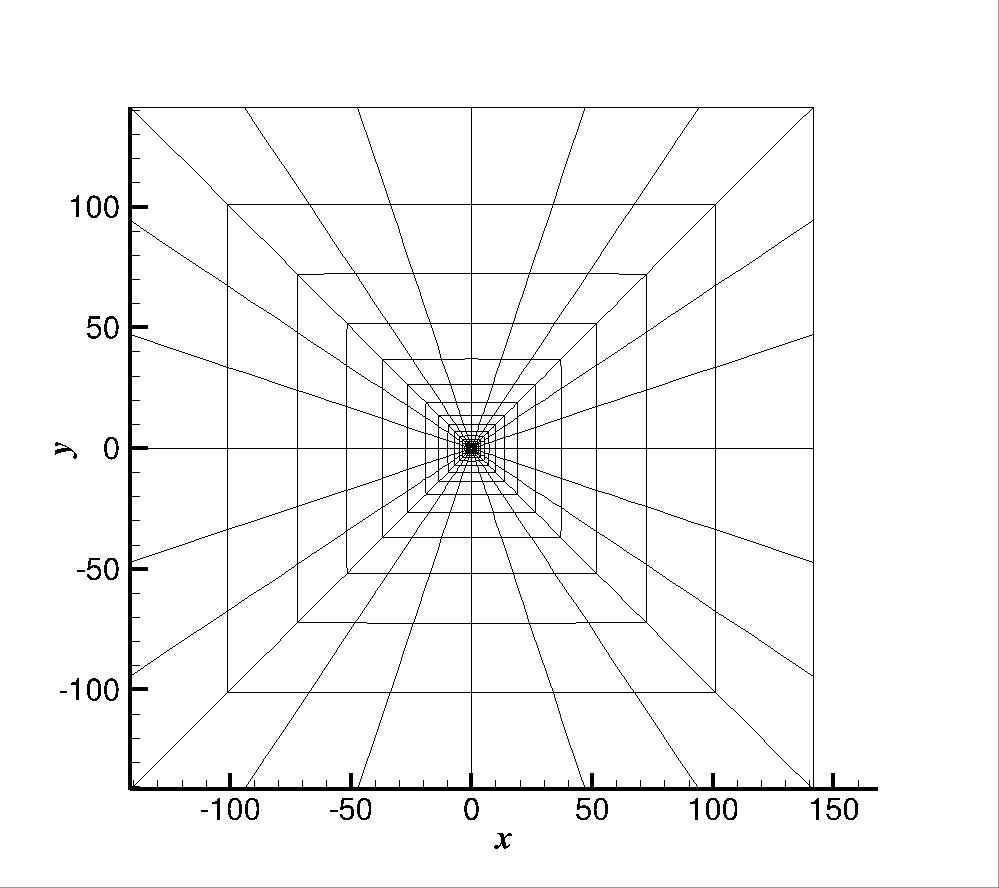} &
	\includegraphics[width=6.5cm]{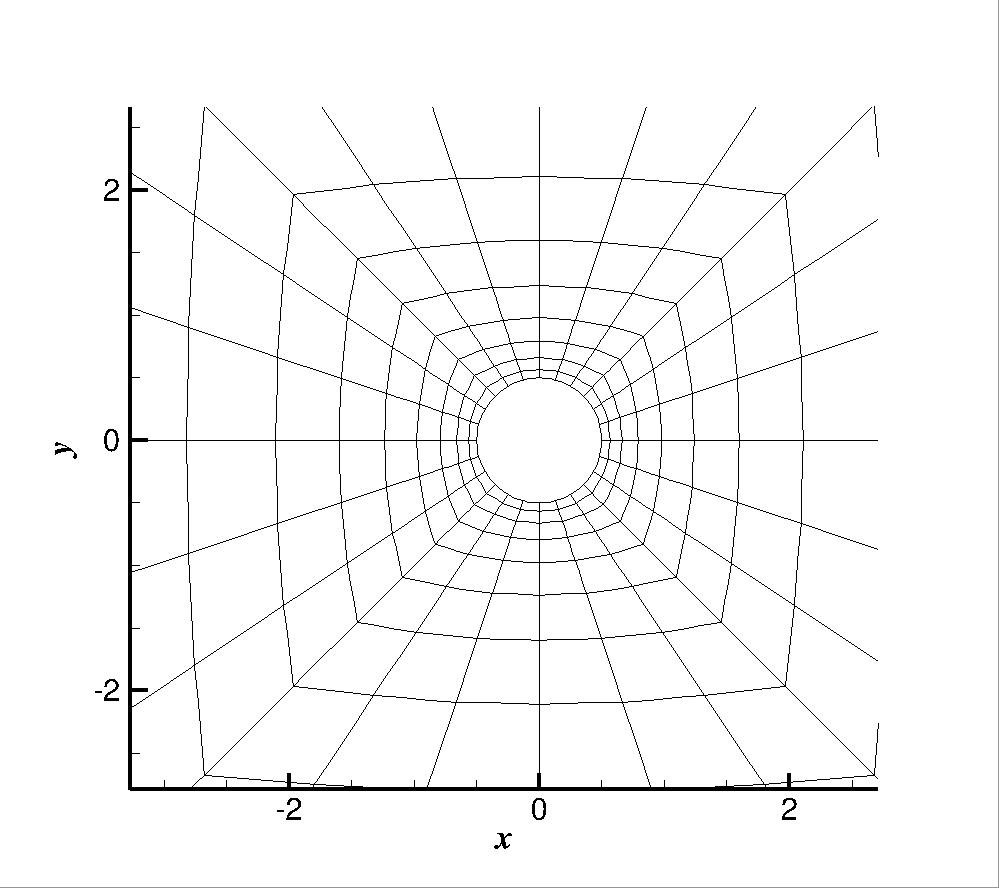}\\
	(a)& (b)\\
	\end{tabular}
	\caption{ 
	(a) An overview of the computational domain of the baseline $ 20\times24 $ mesh, and (b) an enlarged view of the mesh close to the cylinder.}
	\label{cylinder_mesh}
	\end{figure}

	The solver developed in the present study is tested for the inviscid flow over a circular cylinder (smooth curvature) in this section. The physical domain of the problem is within a square of width $ 200\sqrt{2} $. The diameter of the circular cylinder is one. 
	The baseline mesh used in this section is illustrated in Figure~\ref{cylinder_mesh}, which has 20 elements in the normal direction of the wall and 24 elements in the circumferential direction. On splitting the baseline mesh once and twice, one can obtain the $ 40\times48 $ and $ 80\times 96 $ meshes, respectively.

\begin{figure}[t]
	\centering
	\begin{tabular}{cc}
			\includegraphics[width=6.5cm]{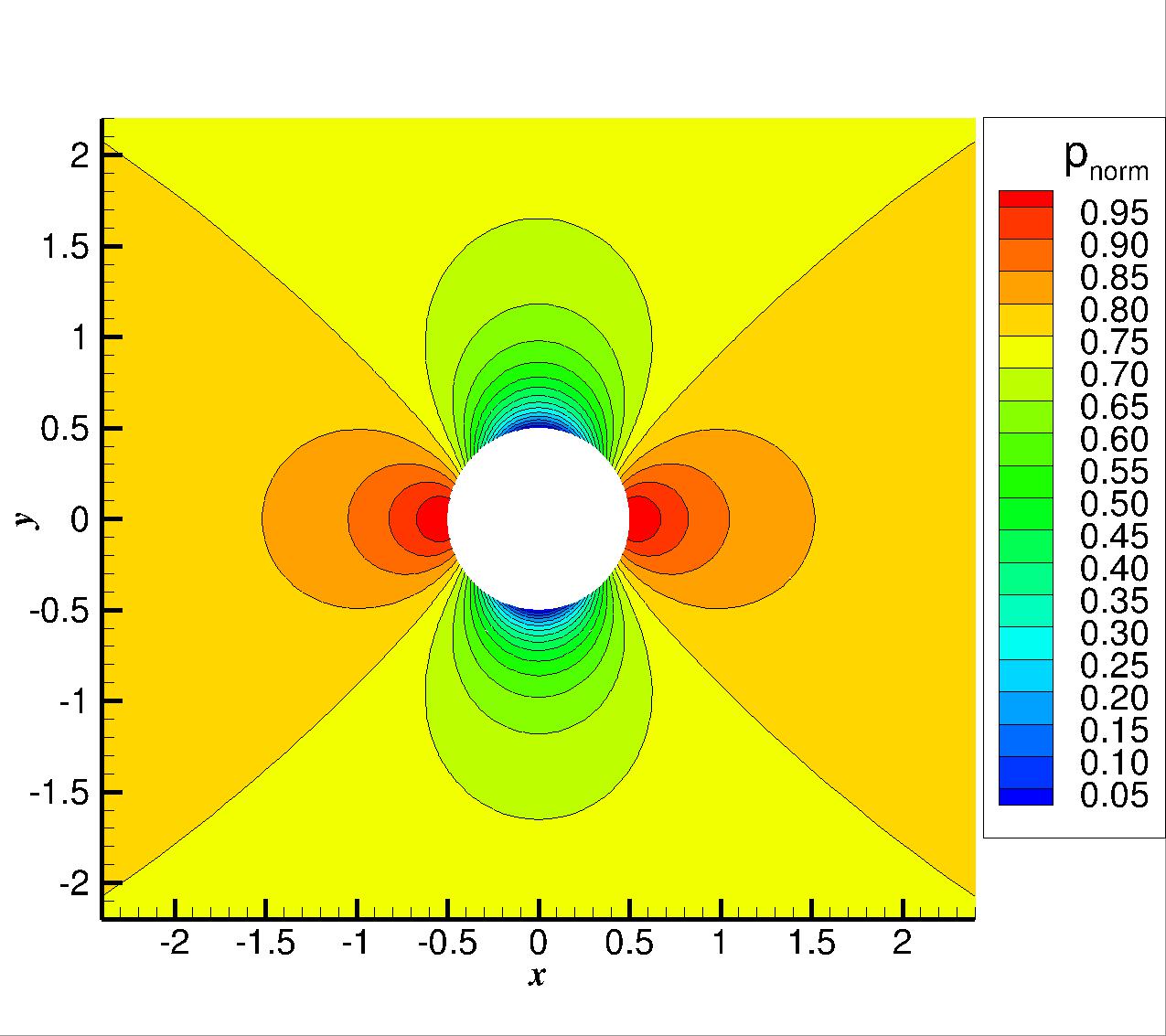} &
			\includegraphics[width=6.5cm]{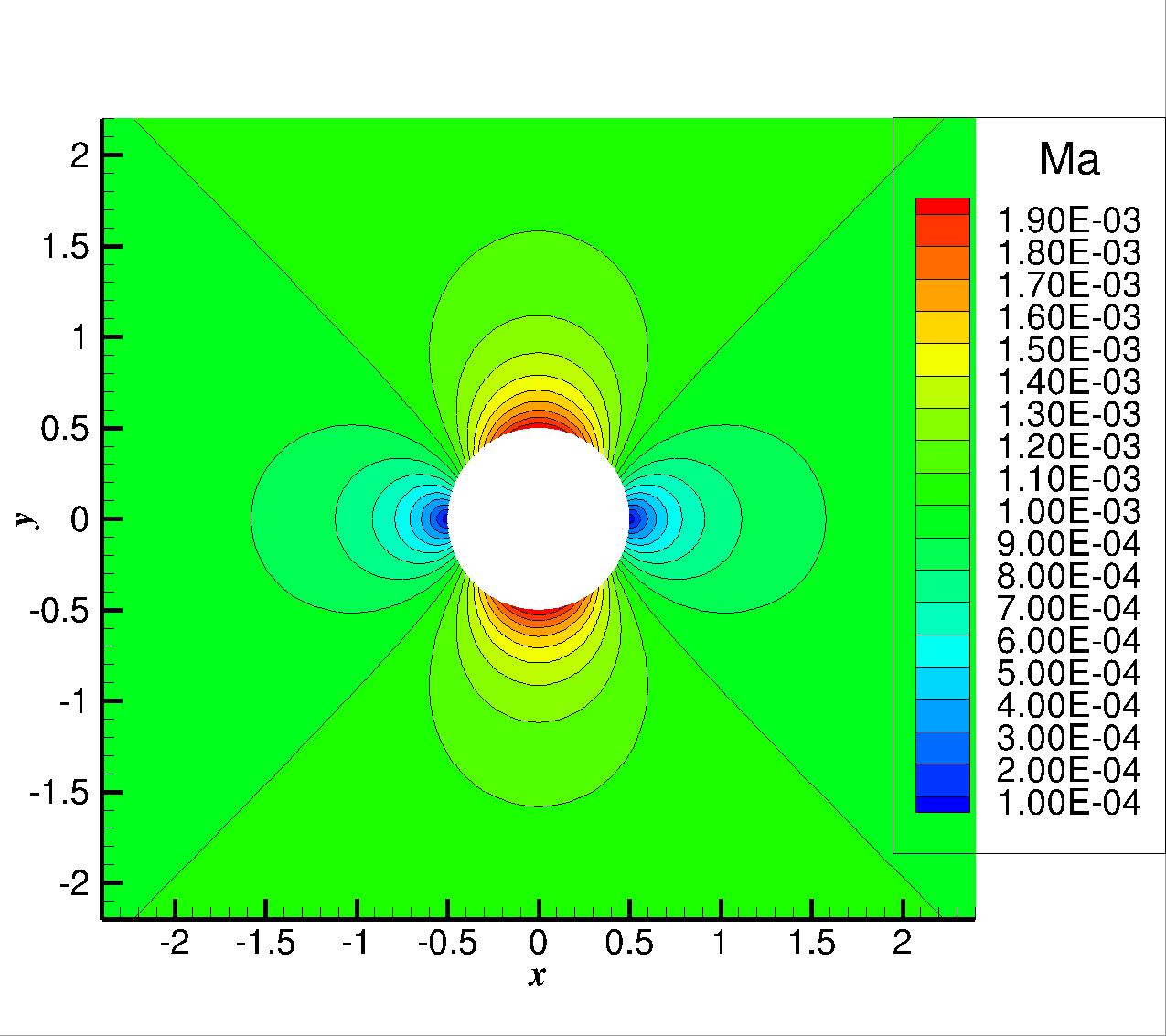}\\
			(a)& (b)\\
	\end{tabular}
	\caption{The inviscid flow over a circular cylinder at $ Ma_\infty=10^{-3} $. (a) $ p_{norm} $ and (b) $ Ma $ contours of the third-order FR/CPR scheme on the $ 80\times96 $ mesh. The inviscid free stream Mach number $Ma_{\infty}$ is $10^{-3}$.}
	\label{cylinder_flowfield}
	\end{figure}

	Grid refinements for the third- ($ P^2 $) and fourth-order ($ P^3 $) FR/CPR schemes have been conducted. The free parameter $\kappa = 1 $ is employed. The convergence criterion for the relative residual of the restarted GMRES solver is  $ tol_{res}={10^{-6}} $ as suggested in Ref.~\cite{Bassi2009}. 
	In Figure~\ref{cylinder_flowfield}, the pressure and $ Ma $ contours near the cylinder in a free stream with $Ma_{\infty}=10^{-3}$ are presented. No pressure oscillation near the stagnation point is observed. Table~\ref{cylinder_grid_refinement} documents the drag coefficient $ C_d $ and entropy error $ E_s $ from a grid refinement study.

		\begin{figure}[]
		\centering
		\begin{tabular}{cc}
		\includegraphics[width=6.5cm]{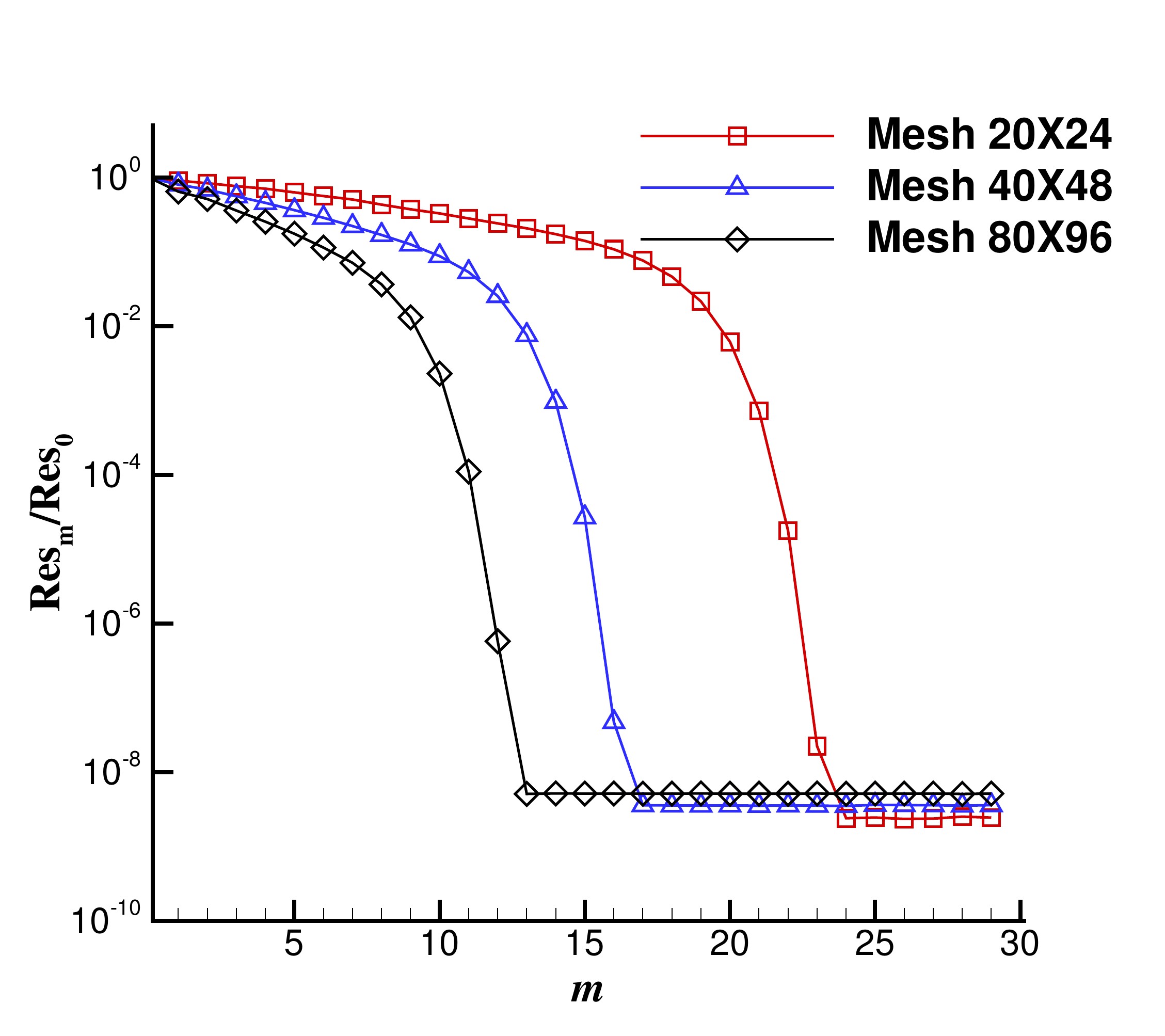} &
		\includegraphics[width=6.5cm]{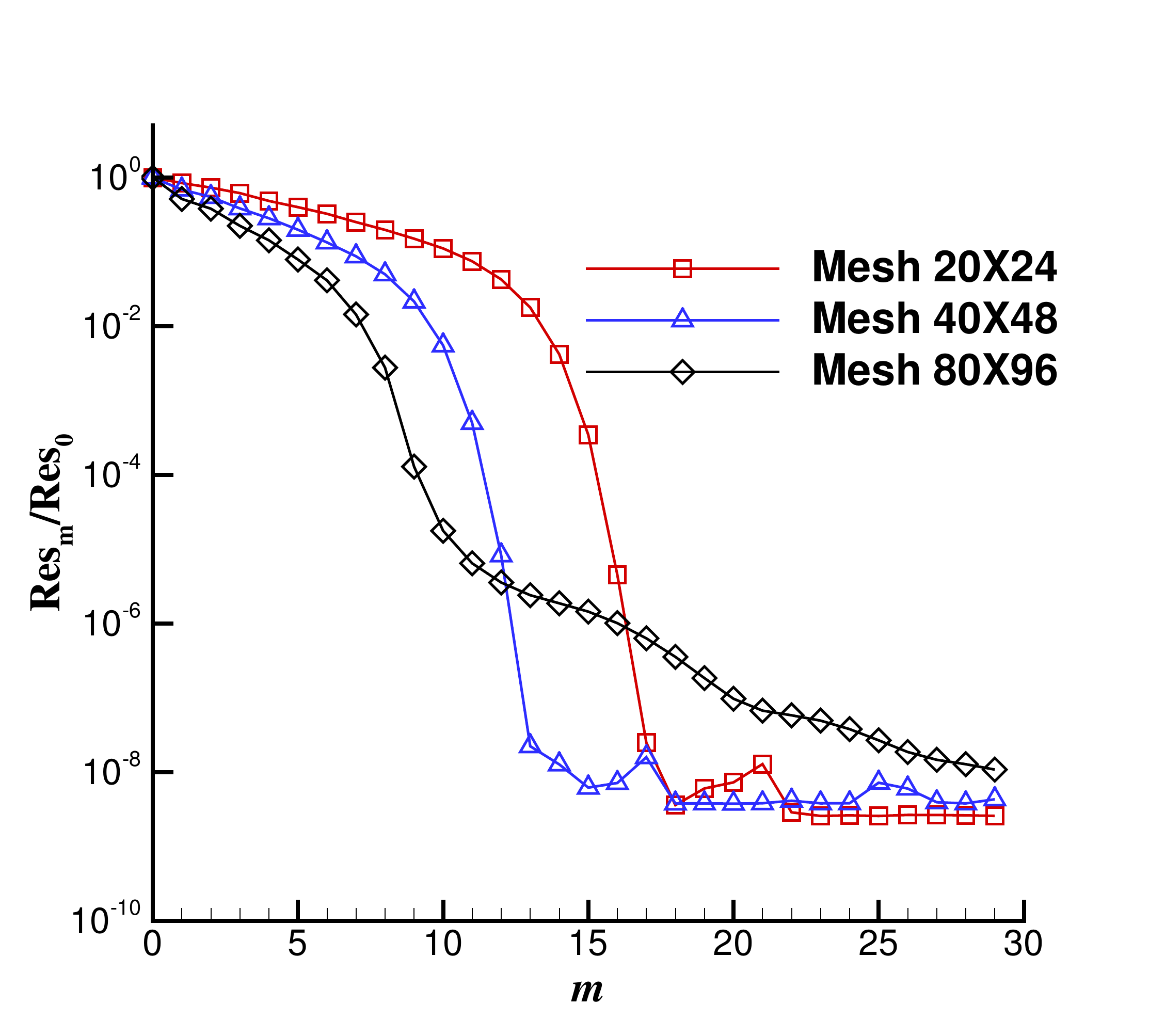}\\
		(a)& (b)\\
		\end{tabular}
		\caption{Convergence histories of $ Res_{m}/Res_{0} $ of pressures for the (a) third-order and (b) fourth-order FR/CPR schemes on different meshes from the simulation of inviscid flow over a circular cylinder at $ Ma_{\infty}=10^{-3} $.}
		\label{cylinder_relres}
		\end{figure}
		
		\begin{figure}[]
			\centering
			\begin{tabular}{cc}
				\includegraphics[width=6.5cm]{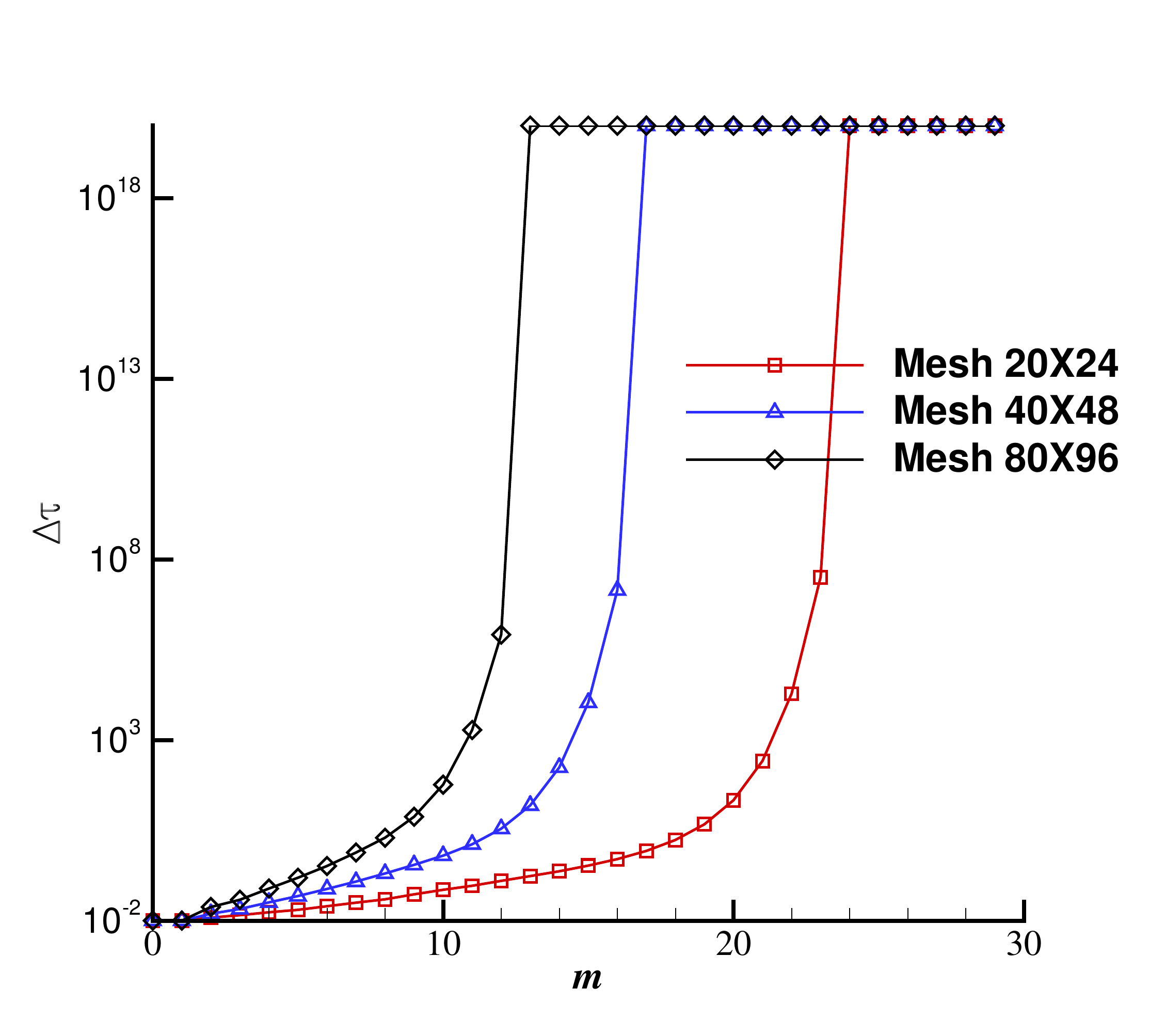} &
				\includegraphics[width=6.5cm]{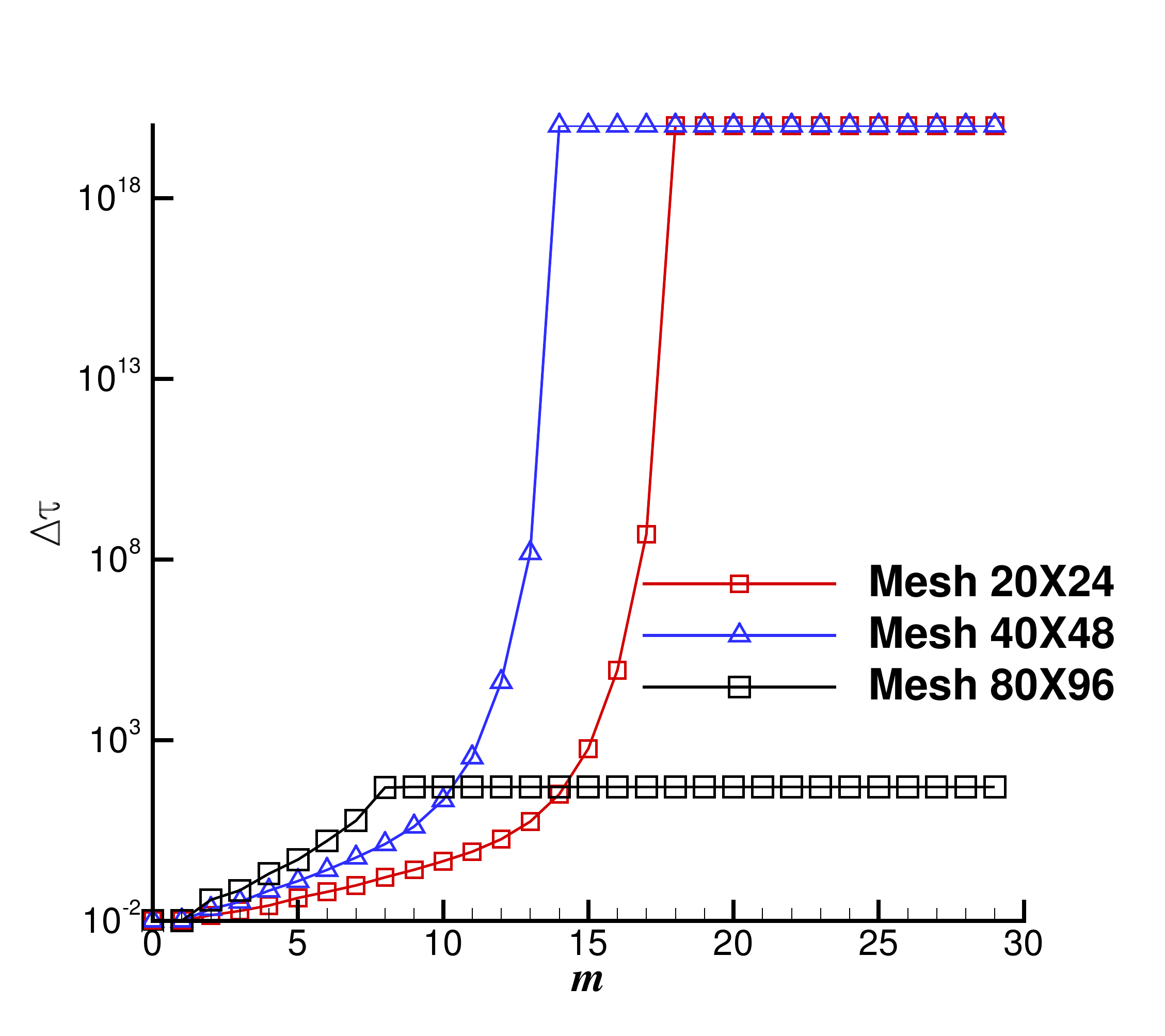}\\
				(a)& (b)\\
			\end{tabular}
			\caption{Pseudo time step $ \Delta \tau $ histories of the pseudo-time iteration for the restarted GMRES solver on different meshes when the (a) third-order and (b) fourth-order FR/CPR schemes are used in the simulation of inviscid flow over a circular cylinder at $ Ma_{\infty}=10^{-3} $.}
			\label{cylinder_dtau}
		\end{figure}
		
		\begin{figure}[]
		\centering
		\begin{tabular}{cc}
			\includegraphics[width=6.5cm]{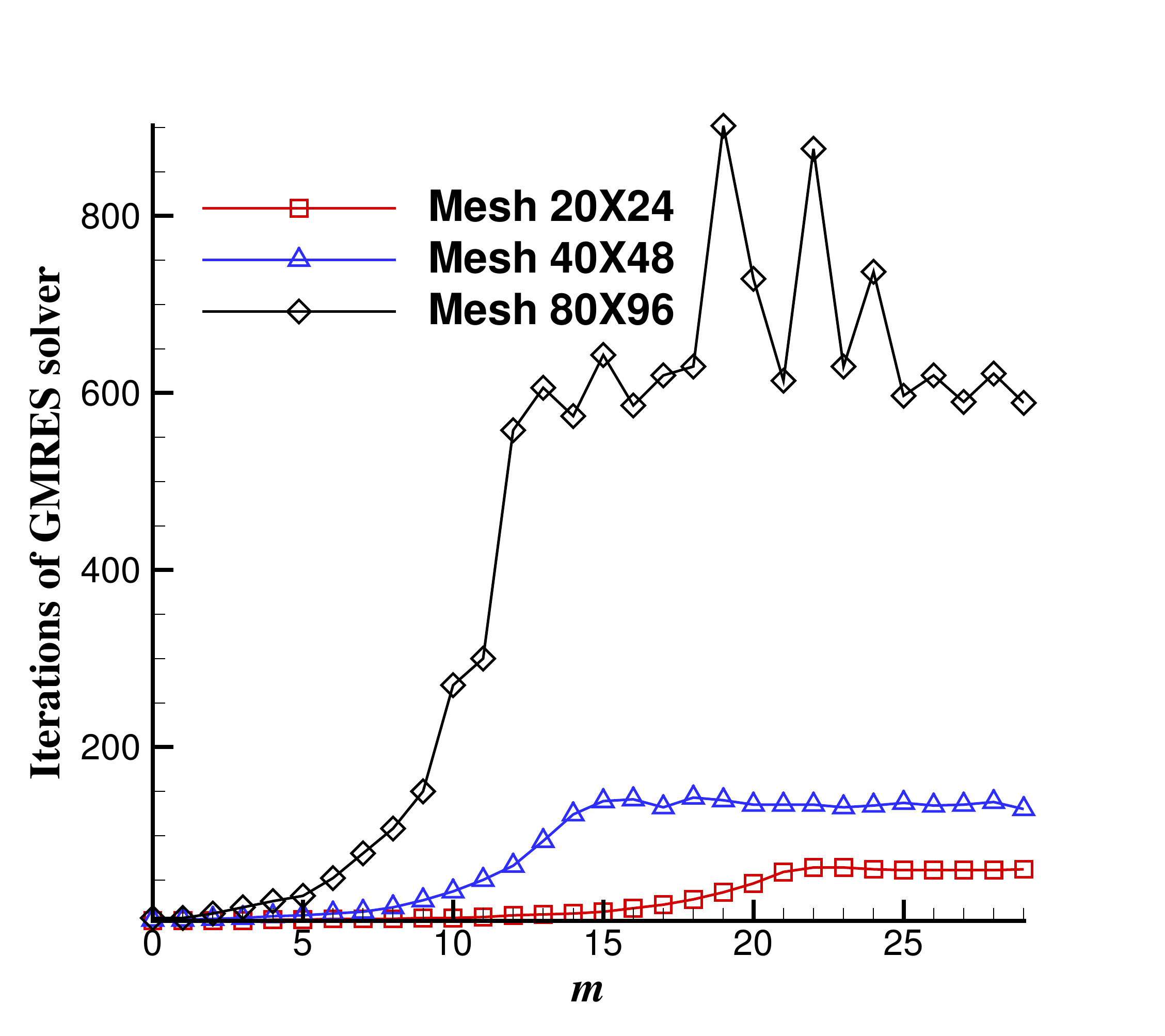} &
			\includegraphics[width=6.5cm]{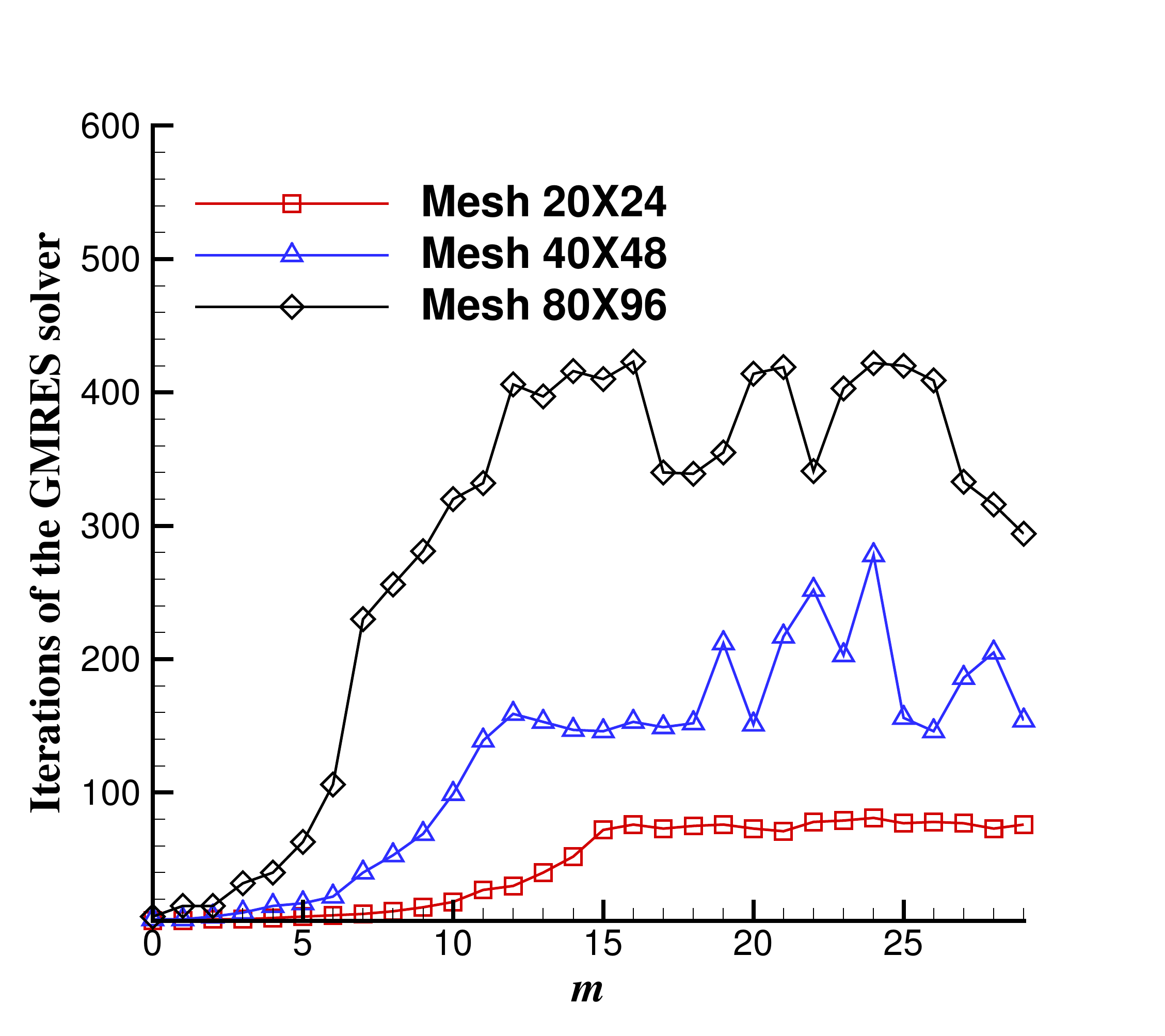}\\
			(a)& (b)\\
		\end{tabular}
		\caption{Iteration numbers of the restarted GMRES solver for the (a) third-order and (b) fourth-order FR/CPR schemes on different meshes from the simulation of inviscid flow over a circular cylinder at $ Ma_{\infty}=10^{-3} $.}
		\label{cylinder_iterations}
		\end{figure}
	
	\begin{table}[]
	\centering
	\caption{The grid refinement study for the inviscid flow over a circular cylinder at $ Ma_\infty=10^{-3} $ with $ tol_{res} = 10^{-6} $. }
	\label{cylinder_grid_refinement}
	\begin{tabular}{cllll}
	\hline
	& \multicolumn{2}{c}{$3^{rd}$} & \multicolumn{2}{c}{$4^{th}$} \\
	\hline
	Mesh & $ C_d $&$ E_s $&$ C_d $&$ E_s $    \\
	\hline
	$20\times24$&$5.2720\times 10^{-4}$&$ 7.0302\times 10^{-12}$  & $ 9.3369\times 10^{-5} $&$ 2.6179\times 10^{-13} $            \\
	$40\times48$&$1.7897\times 10^{-4}$&$ 6.9203\times 10^{-13}$  & $ 6.7059\times 10^{-6} $&  $ 2.0882\times10^{-14} $      \\
	$80\times96$&$2.8318\times 10^{-5}$&$ 7.3010\times 10^{-14}$  & $ 4.5020\times10^{-7} $ & $ 3.2653\times10^{-15} $            \\
	\hline   
	\end{tabular}
	\end{table}
	
	The convergence histories of the relative residual $  Res_m/Res_0 $ of pressures are illustrated in Figure~\ref{cylinder_relres}. It is observed that $ Res_m/Res_0 $ can decrease to the level of $ 10^{-9} $ in steady flow simulation at low Mach numbers. The histories of the pseudo-time step size variation are presented in Figure~\ref{cylinder_dtau}. An observation is that $  Res_m/Res_0 $ will quickly drop as the pseudo-time step size increases to large values. This is due to that the Newton's method will be recovered as $ \Delta \tau \to \infty$. However, as shown in Figure~\ref{cylinder_iterations}, the iteration numbers needed for the restarted GMRES solver to converge (i.e., to meet the convergence criterion $ tol_{res}={10^{-6}} $) will significantly increase when $ \Delta \tau \to \infty $, especially when fine meshes are used. A remedy could be decreasing the maximum value of $ \Delta \tau $. As presented in Figure~\ref{cylinder_iterations}(b), when solving the problem with the fourth-order FR method on the $ 80\times96 $ mesh, $ \Delta \tau_{max} = 50 $ is employed to ensure that the iteration number needed by the GMRES solver to converge is around 400 when $ \Delta \tau = \Delta \tau_{max} $. Consequently, more pseudo-time iterations are needed for the pseudo transient procedure to converge as shown in Figure~\ref{cylinder_relres}(b).
	
	\subsubsection*{Effects of the Global Cut-off Parameter $ \kappa $}
\begin{figure}[]
	\centering
	\begin{tabular}{cc}
	\includegraphics[width=7cm]{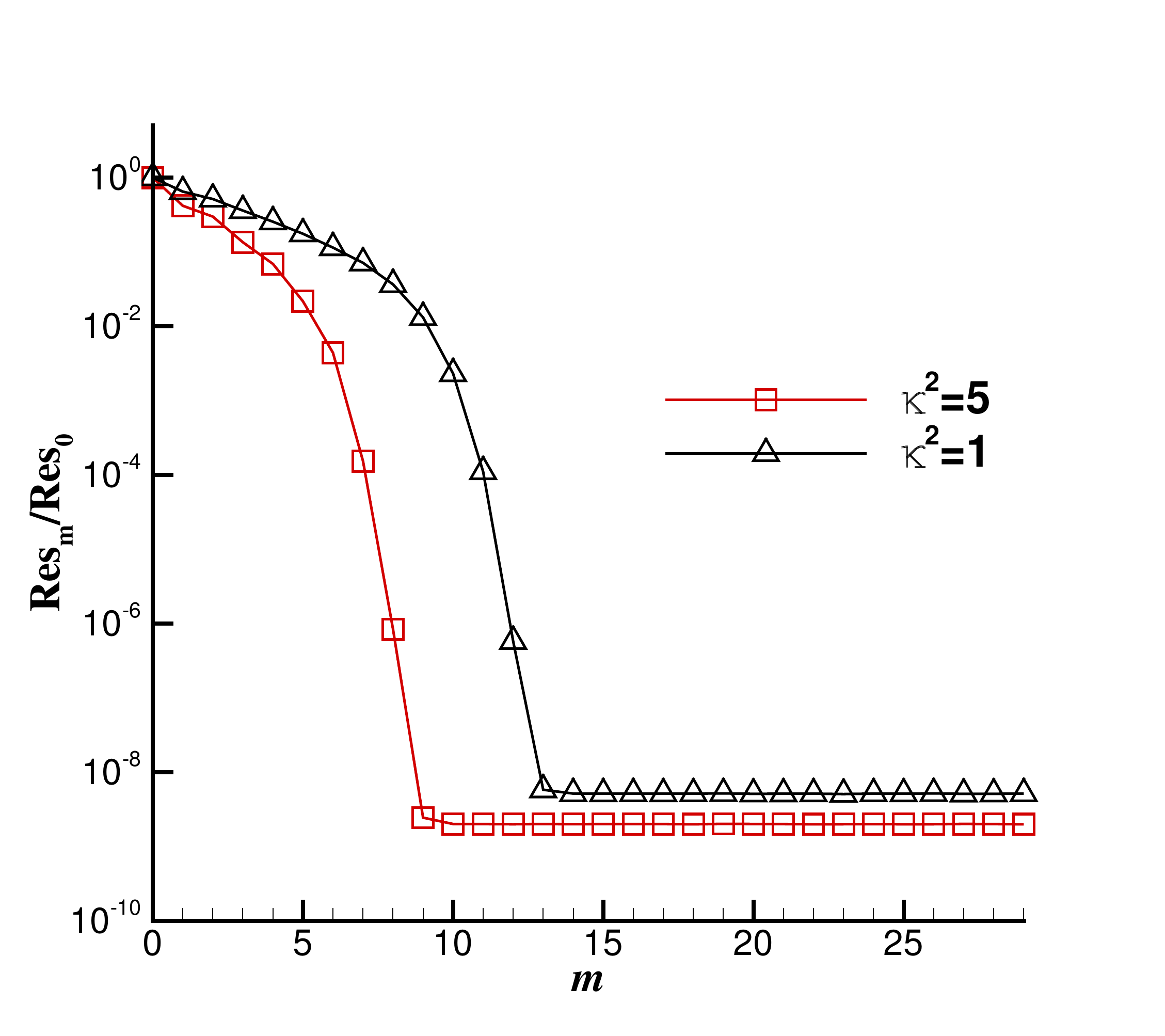} &
	\includegraphics[width=7cm]{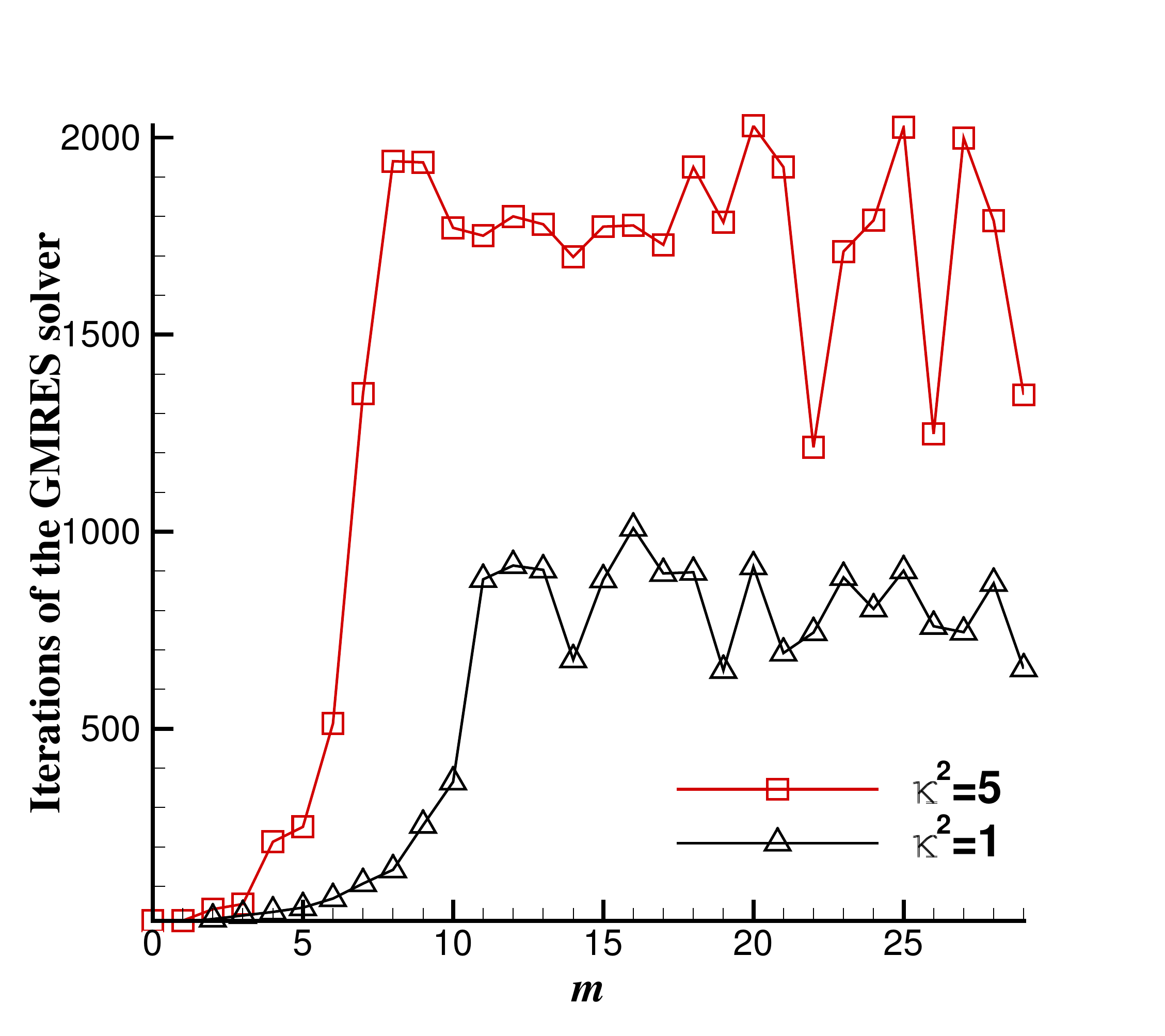}\\
	(a)& (b)\\
	\end{tabular}
	\caption{Histories of (a) $ Res_m/Res_0 $, and (b) the corresponding iteration numbers of the restarted GMRES solver for different $ \kappa '$s from the simulation of inviscid flow over a circular cylinder at $ Ma_{\infty}=10^{-3} $.}
	\label{iters_res_diff_kappa}
	\end{figure}

	We have tested three different values of $ \kappa $, namely, $ \kappa = \sqrt{0.2} $, $ \kappa=1 $ and $ \kappa = \sqrt{5} $ to study the effects of $ \kappa $ on convergence and accuracy. The convergence criterion for the relative residual of the restarted GMRES solver is set as  $ tol_{res}={10^{-9}} $ in these tests. For brevity, only the third-order FR/CPR scheme is considered on the $ 80\times 96  $ mesh. The results of $ C_d $, $ C_l $ and $ E_s $ are presented in Table~\ref{effect_of_kappa}. 
	The histories of relative residuals and the iteration numbers needed for each pseudo-time stepping are presented in Figure~\ref{iters_res_diff_kappa}.
	When $ \kappa = \sqrt{0.2} $, the global cut-off value is very small. Due to the instability induced by the two stagnation points, the linear solver is not able to converge within the maximum iteration number (5000 for this case), and the simulation quickly diverged. Even though when $ \kappa = \sqrt{5} $, $ Res_m/Res_0 $ can drop to a smaller value using less pseudo-time iterations compared to that with $ \kappa = 1 $, the accuracy does not get better, and more iterations are needed for the linear solver to converge in one pseudo-time step. The predicted $ C_d $ and $ E_s $ both increase due to the fact that when a larger $ \kappa $ is used, the maximum absolute eigenvalue $ |\lambda|_{max} $ in Eq.~\eqref{eigenvalues} is increased. Therefore, more numerical dissipation is added to the approximate Riemann solver.
	\begin{table}[]
		\centering
		\caption{$ C_l $, $ C_d $ and $ E_s $ for different $ \kappa '$s when the third-order FR/CPR scheme is used to simulate the inviscid flow over a circular cylinder at $ Ma_\infty=10^{-3} $ on the $ 80\times96 $ mesh with $ tol_{res} = 10^{-9} $.} 
		\label{effect_of_kappa}
		\begin{tabular}{clll}
			\hline
			\multicolumn{1}{l}{$ \kappa $}    & $ C_d $       & $ C_l $  & $ E_s $      \\
			\hline
			$ \sqrt{0.2}  $  & diverged      & diverged &  diverged                     \\
			$1$& $ 2.8319\times 10^{-5}  $& $ 6.6839\times 10^{-7}  $       & $ 7.3013\times 10^{-14} $                  \\
			$\sqrt{5}$& $ 5.1692\times10^{-5}  $  & $  1.2015\times 10^{-6}$         &  $ 1.3620\times 10^{-13}  $             \\
			\hline   
		\end{tabular}
	\end{table}
	
	\subsection{Inviscid Flow over a NACA0012 Airfoil} \label{Steady_NACA12}
\begin{figure}[]
	\centering
	\begin{tabular}{cc}
		\includegraphics[width=6cm]{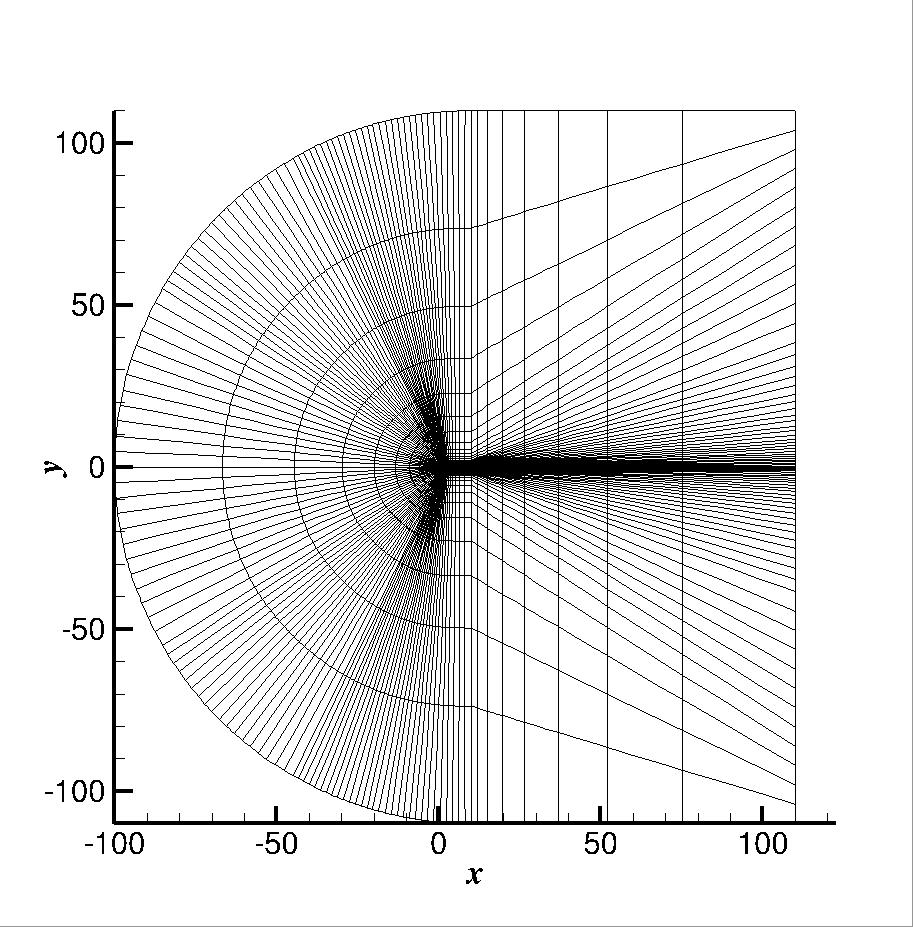} &
		\includegraphics[width=6cm]{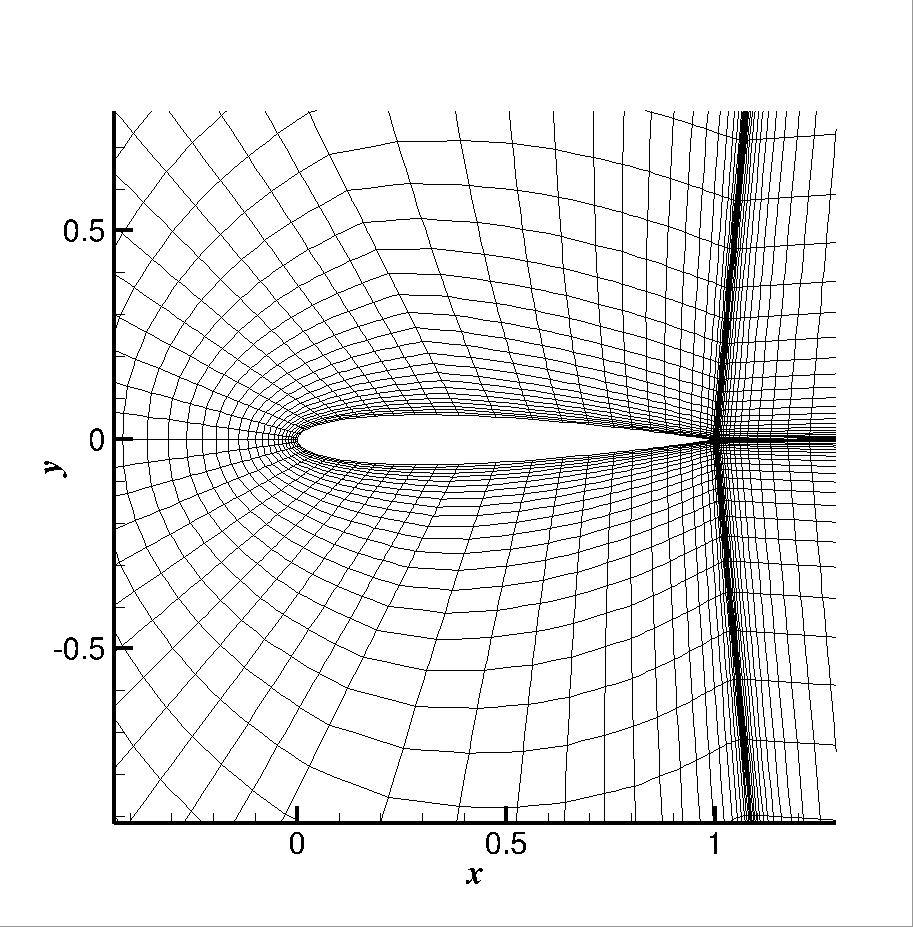}\\
		(a)& (b)\\
	\end{tabular}
	\caption{(a) Domain and mesh of the flow over a NACA0012 airfoil, and (b) an enlarged view of the mesh close to the airfoil.}
	\label{naca_mesh}
	\end{figure}

\begin{figure}[]
		\centering
		\begin{tabular}{cc}
			\includegraphics[height=6cm]{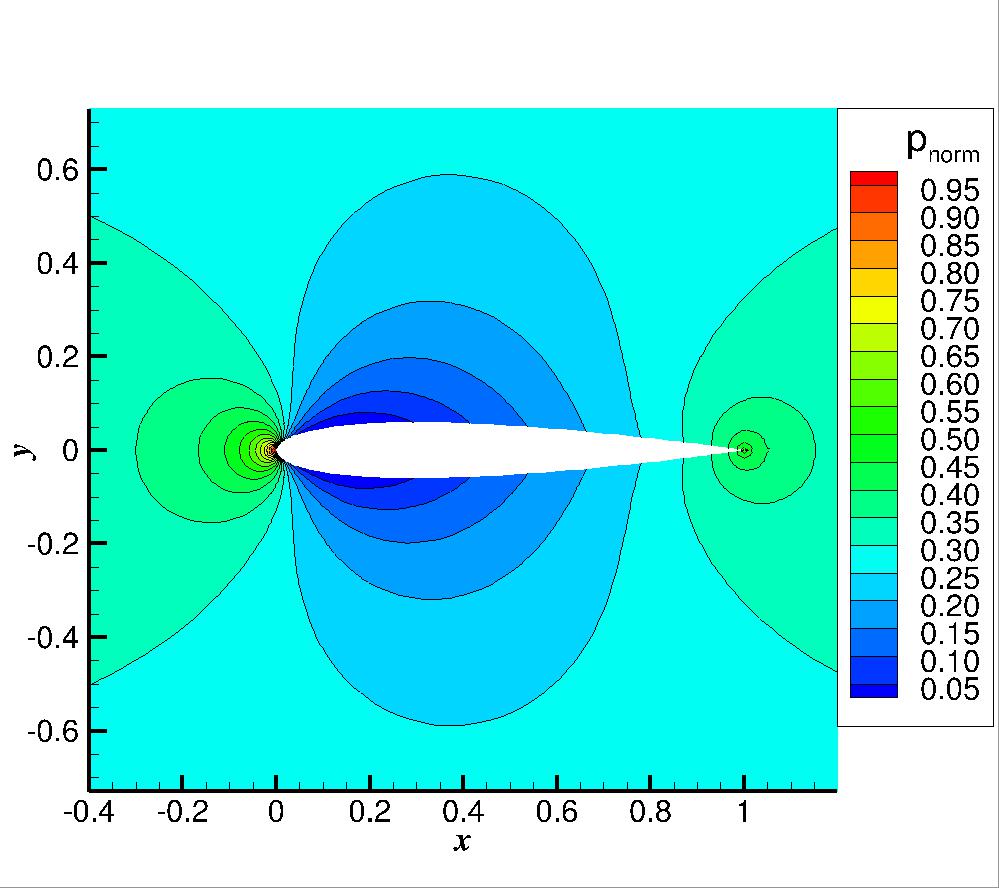} &
			\includegraphics[height=6cm]{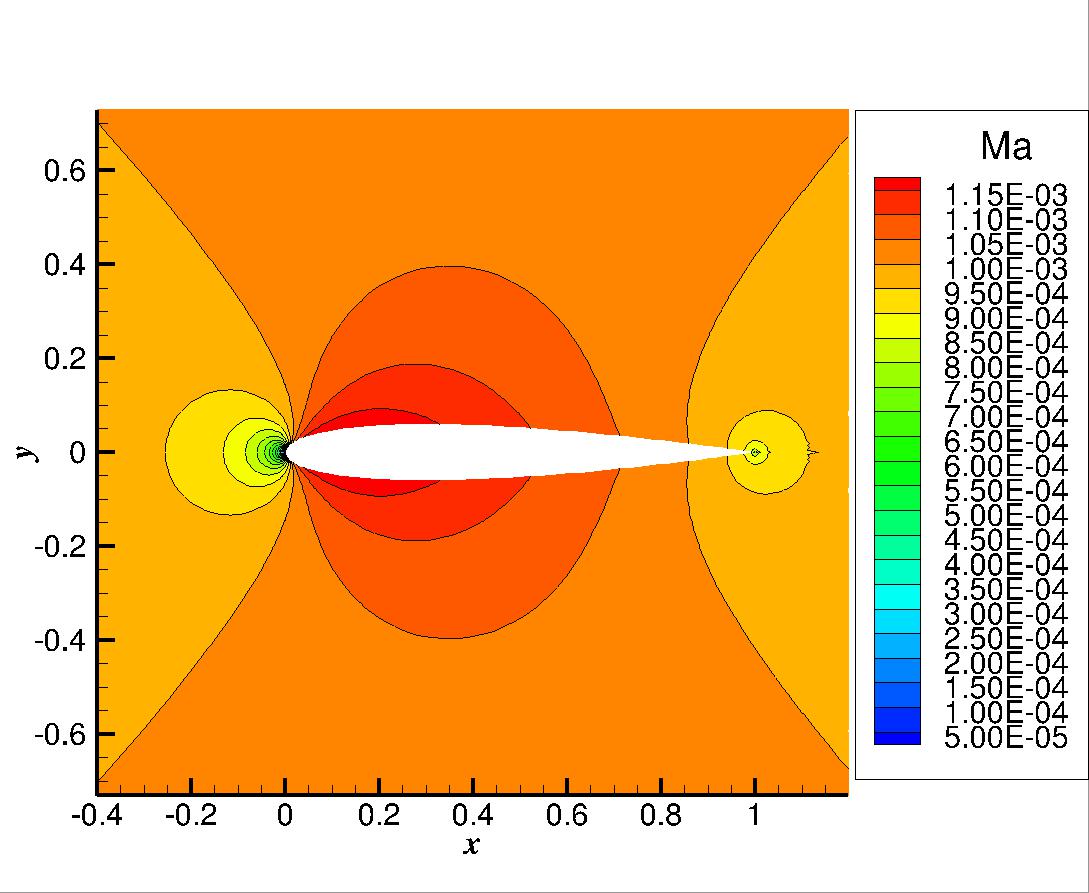}\\
			(a)& (b)\\
		\end{tabular}
		\caption{(a) $ p_{norm} $ and (b) $ Ma $ contours of the flow over a NACA0012 airfoil using the third-order FR/CPR method with $ tol_{res} = 10^{-6} $. The free stream Mach number $Ma_{\infty}$ is $10^{-3}$.}
		\label{naca_flow}
	\end{figure}

	In this section, the inviscid flow of $ Ma=10^{-3} $ and $ 0^\circ $ angle of attack (AOA) over the NACA0012 airfoil is  studied to verify the performance of the preconditioned FR/CPR solver when there exists a singular point on the geometry. The profile of the airfoil is expressed as 
	\begin{equation}
	y=\pm0.6\left(0.2969\sqrt{x}-0.1260x-0.3516x^2+0.2843x^3-0.1036x^4\right),
	\end{equation}
	where $ x\in[0,1] $.
	The mesh with 5,168 quadrilateral elements is illustrated in Figure~\ref{naca_mesh}.
	The polynomial degree is refined from $ 2 $ to $ 4$ (i.e., from the $3^{rd}$ to $5^{th}$ order of accuracy). As presented in Table~\ref{naca_refinement},  the prediction of $ C_d $  becomes more accurate when the degree of the polynomial increases. With preconditioning, instabilities near wall boundaries are suppressed and the flow field in Figure~\ref{naca_flow} does not show any pressure oscillation near wall boundaries. We have also tested the impact of different criteria $ tol_{res} $, i.e., $ 10^{-3} $, $ 10^{-6} $ and $ 10^{-9} $, on the convergence.
	
\begin{figure}[]
	\centering
	\begin{tabular}{c}
	\includegraphics[width=12cm]{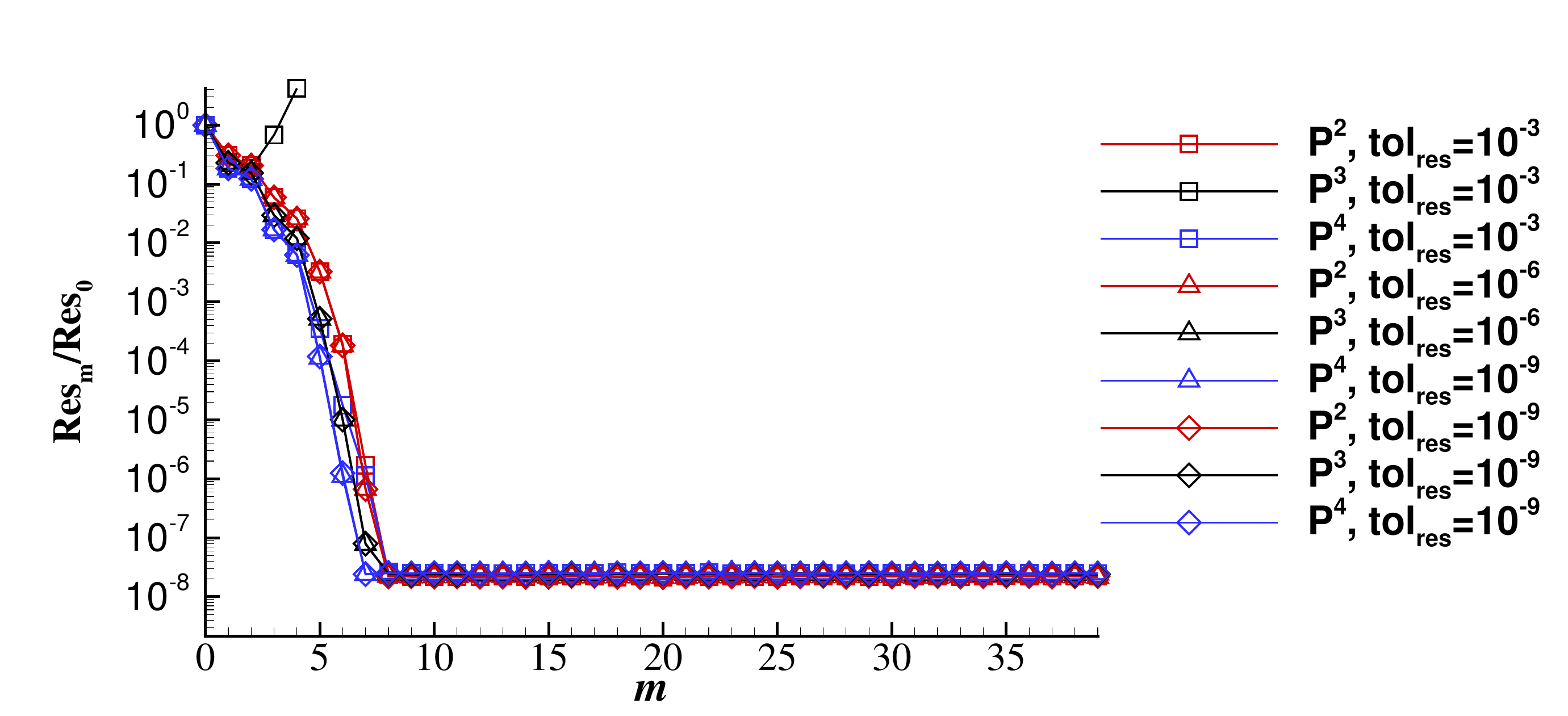} \\
	(a)\\
	\includegraphics[width=12cm]{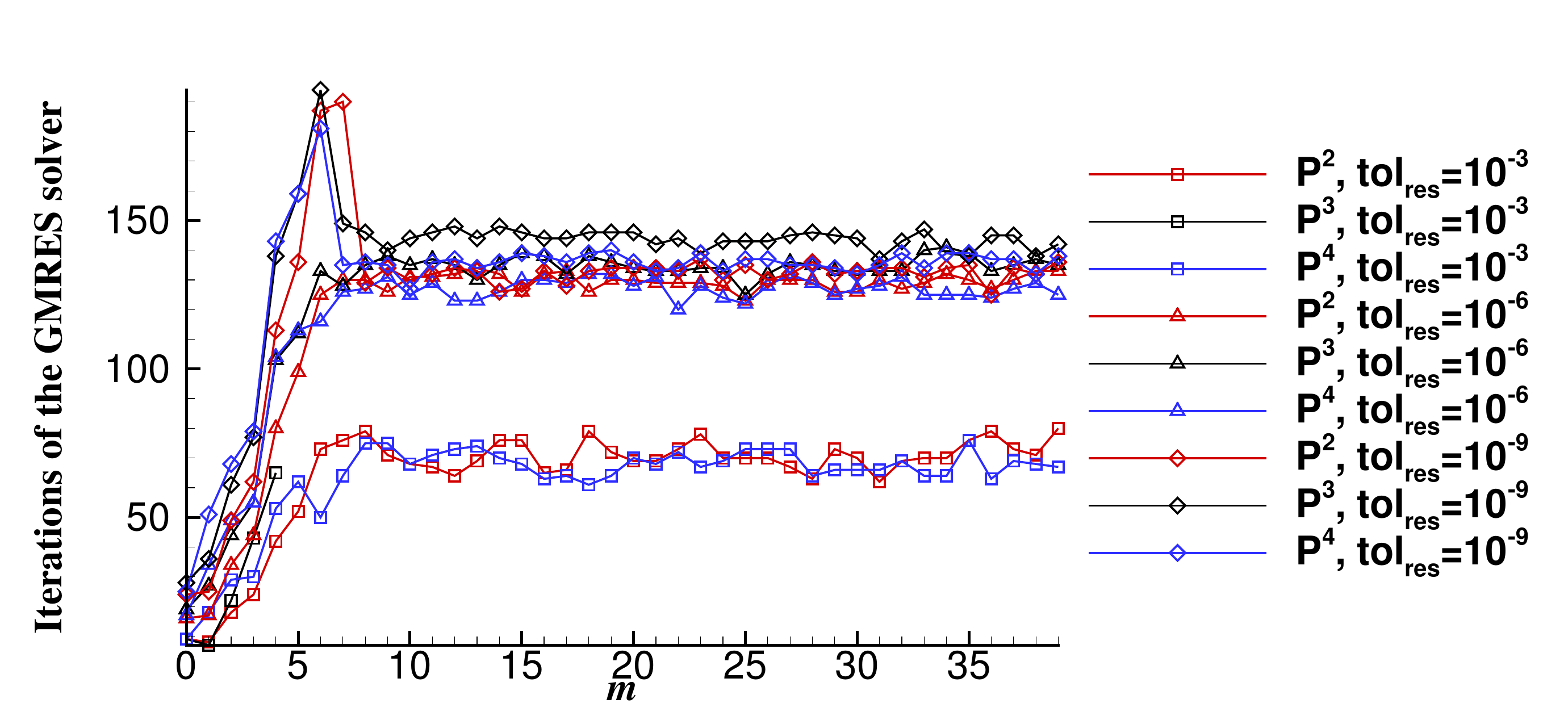}\\
	(b)\\
	\end{tabular}
	\caption{Histories of (a) $ Res_m/Res_0 $ and (b) iteration numbers of the restarted GMRES solver of the third-, fourth-, and fifth-order FR/CPR schemes with $ tol_{res}=10^{-3} $, $ 10^{-6} $ and $ 10^{-9} $ for solving the inviscid flow over a NACA0012 airfoil at $ Ma_\infty=10^{-3} $.}
	\label{naca_res_iterations}
	\end{figure}

	In Ref.~\cite{Bassi2009}, a 1,792-element mesh is employed and the maximum iteration number of the restarted GMRES solver when $ \Delta\tau\to\infty $ is roughly 120. In this study, a relative denser mesh is used. We observe from Figure~\ref{naca_res_iterations}(b) that the maximum iteration number is about $ 140 $ with the same $ tol_{res}=10^{-6} $ as that used in Ref.~\cite{Bassi2009}. According to the convergence comparison of different criteria for the restarted GMRES solver (see Figure~\ref{naca_res_iterations}(a)), when $ tol_{res} = 10^{-3}$, both the third-order and fifth-order FR/CPR schemes can converge to correct solutions. However, the simulation using the fourth-order FR/CPR scheme will immediately diverge after a few pseudo-time iterations. With the SER approach employed in this study, a smaller $ tol_{res} $ for the linear solver is suggested for the sake of convergence, but more iterations of the restarted GMRES solver are needed (see Figure~\ref{naca_res_iterations}(b)). Only negligible differences are observed on $ C_d $ and $ E_s $ among the converged results of different $ tol'_{res} $s, which are not presented for brevity. Therefore, a proper choice of $ tol_{res} $ is a trade-off between robustness and efficiency.
	
	\begin{table}
	\centering
	\caption{The order of accuracy refinement study for the inviscid flow over the NACA0012 airfoil at $ Ma_{\infty}=10^{-3} $ with $ tol_{res} = 10^{-6} $.} 
	\label{naca_refinement}
	\begin{tabular}{clll}
	\hline
	& $3^{rd}$& $4^{th}$ &$5^{th}$\\
	\hline
	$ C_d $&$7.5664\times 10^{-6}$ &$ 2.7816\times 10^{-6}$  & $ 2.1640\times10^{-6} $       \\
	$ E_s $&$2.4161\times 10^{-12}$&$ 4.0361\times 10^{-13}$ & $ 1.8262\times10^{-13} $  \\	
	\hline   
	\end{tabular}
	\end{table}
	
	\subsection{Isentropic Vortex Propagation with Prescribed Grid Motion}
	The isentropic vortex propagation case is employed to verify both the spatial convergence and temporal convergence of the preconditioned FR/CPR methods for unsteady flow simulation in this section.
	The isentropic vortex propagation case depicts the superposition of an inviscid uniform flow and an irrotational vortex. The vortex can be regarded as a perturbation added onto the uniform flow. The free stream flow is of $ (\rho,u,v,Ma) = (1.0,\sqrt{2}/2,\sqrt{2}/2,0.05) $ and the gas constant $ R=1.0 $ for this case. The perturbation is defined as~\cite{Bassi2015}
	\begin{equation}\label{VP_Solution}
	\begin{cases}
	\delta u=-\frac{\alpha}{2\pi}(y-y_0)e^{\phi(1-r^2)}\\
	\delta v=\frac{\alpha}{2\pi}(x-x_0)e^{\phi(1-r^2)}\\
	\delta T=-\frac{\alpha^2(\gamma-1)}{16\phi\gamma\pi^2}e^{2\phi(1-r^2)}\\
	\end{cases}	
	\end{equation}
	where $ \phi = \frac{1}{2}$ and $ \alpha = 5 $ are parameters that define the vortex strength. $ r=(x-x_0)^{2}+(y-y_0)^{2} $ is the distance to the center of the vortex $ (x_0,y_0) $. The vortex is propagated in a periodic domain $ [-10,10]\times[
	-10,10] $. A prescribed grid motion is employed to validate the moving grid algorithm. The prescribed grid motion is defined as 
	\begin{equation}
	\begin{cases}
	x(t) = x_r+a_x \sin(2\pi f_nt)\sin(2\pi f_x x_r)\sin(2\pi f_y y_r)\\
	y(t) = y_r+a_y \sin(2\pi f_nt)\sin(2\pi f_x x_r)\sin(2\pi f_y y_r)
	\end{cases}
	\end{equation}
	where $ x_r,y_r $ are the coordinates at $ t=0 $, i.e.,  coordinates of the uniformly divided grid, and $  a_x = 1$, $ a_y=1 $, $ f_n = 1.0 $, $ f_x=0.1 $, $ f_y = 0.1 $.
	Periodic boundary conditions are applied to all boundaries. Since there is no presence of wall boundaries, the global cut-off parameter $ \kappa $  is set as zero for this case.
	
\begin{figure}[]
	\centering
	\includegraphics[width=8cm]{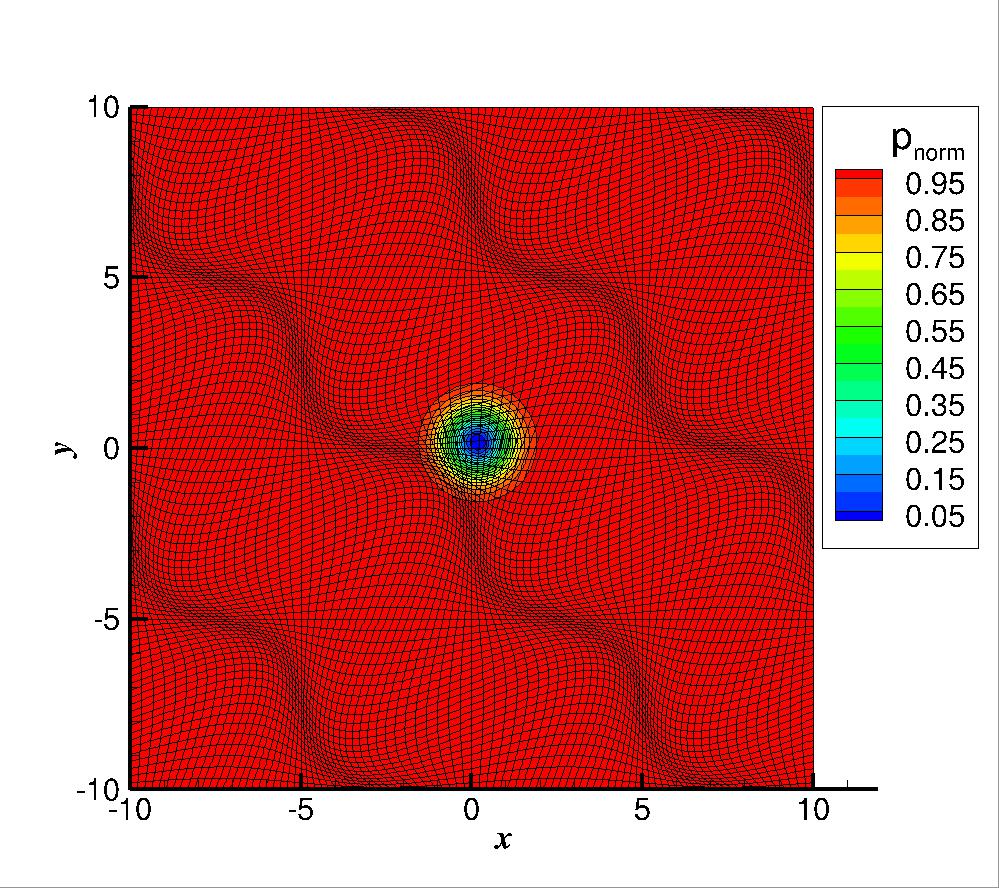}
	\caption{$ p_{norm} $ contour  and the deformed mesh at $ t=\frac{1}{4}T_n $ from the simulation of isentropic vortex propagation with a free stream Mach number $ Ma_\infty =0.05 $. The fifth-order FR/CPR scheme, $ 96\times96 $ mesh, and $ \Delta t =  \frac{1}{800}T_n$ are used in this case.}
	\label{vp_contour}
	\end{figure}

\begin{figure}[]
	\centering
	\begin{tabular}{cc}
	\includegraphics[height=6cm]{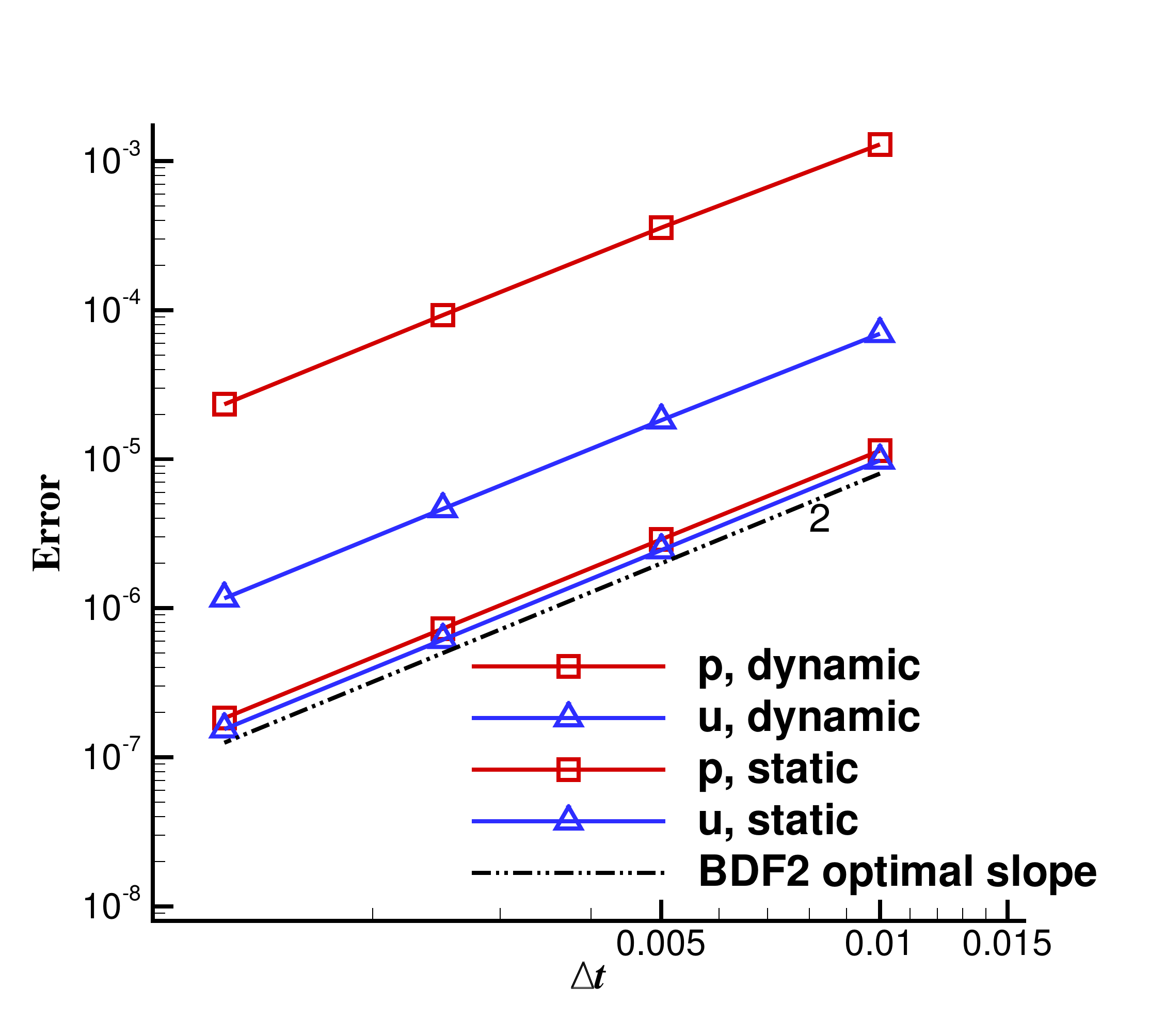} &
	\includegraphics[height=6cm]{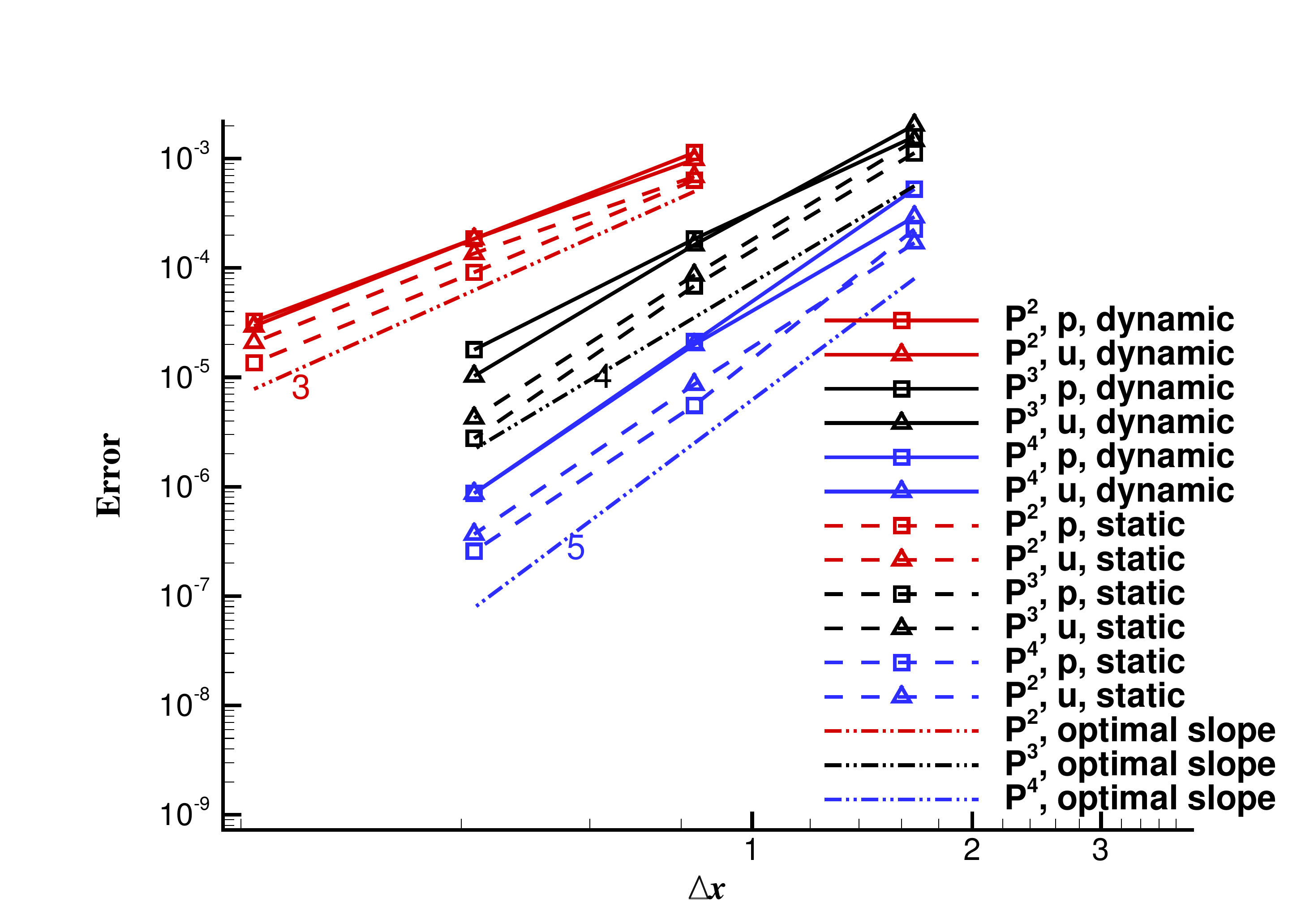}\\
		 (a)&(b)\\
	\end{tabular}
	\caption{(a) The convergence study of BDF2 on the $ 96\times96 $ mesh using the fifth-order FR/CPR scheme, and (b) the grid refinement study of the third-, fourth-, and fifth-order FR/CPR schemes for the isentropic vortex propagation case with a free stream Mach number $ Ma_\infty =0.05 $.}
	\label{vp_refinement}
	\end{figure}

	The time step size $ \Delta t $ is refined from $ T_n/100 $ to $ T_n/800 $, where $ T_n = 1/f_n $ to validate the accuracy of BDF2.  A fifth-order FR/CPR scheme and a $ 96\times96 $ uniformly divided mesh are employed to ensure that the error is dominated by the temporal discretization. We require the relative residual $ Res_{m}/Res_{0} $ of the pseudo-time iterations to drop as low as possible  within 100 pseudo-time iterations. The $ tol_{res} $ of the restarted GMRES solver is $ 10^{-6} $ with the maximum iteration number 100. We only simulate this case until $ \frac{1}{4}T_n $ when the mesh deformation is maximized to demonstrate the convergence of BDF2 with GCL. An illustration of the pressure contour and the mesh deformation can be found in Figure~\ref{vp_contour}. As observed from Figure~\ref{vp_refinement}(a), the optimal convergence rate of BDF2 is achieved for both $ p $ and $ u $. 
	
	A grid refinement study is also conducted on a uniformly-divided mesh set with $12 \times 12$, $24 \times 24$, $48 \times 48 $ and $96 \times 96$ elements, respectively. The time step size $ \Delta t$  is $ 10^{-4} $ to guarantee that the error is dominated by the spatial discretization. The $ L_2 $ error of pressure $ p $ and velocity $ u $ are presented in Figure~\ref{vp_refinement}(b). We observe that for both $ p $ and $u$, the numerical orders of a degree $k$ solution construction are close to $ k+1/2 $ with and without dynamic meshes. Note that the optimal convergence rate is $k+1$. The observed order reduction is reasonable since the Rusanov approximate Riemann solver is employed to calculate the common inviscid fluxes~\cite{Hesthaven08}. Hence, the preconditioned FR/CPR methods can demonstrate good convergence for unsteady low-Mach-number flow simulations on dynamic meshes. 
	\subsection{Laminar Flow over a Plunging NACA0012 Airfoil}
	An order of accuracy refinement study is firstly carried out for a steady case, i.e., the viscous flow of $Re=500$, $ Ma = 10^{-3} $ and $ 0^\circ $ AOA over the stationary NACA0012 airfoil. The mesh is the same as that in Section~\ref{Steady_NACA12} (see Figure~\ref{naca_mesh}). The drag coefficients of the third-, fourth-, and fifth-order FR/CPR schemes are $ 0.1723 $, $ 0.1722 $ and $ 0.1722 $, respectively. Therefore, we adopt the fourth-order FR/CPR scheme to simulate laminar flows over a plunging NACA0012 airfoil. 
	The plunging motion is defined as $ y=Y\sin(2\pi f t) $, where $Y$ is the plunge amplitude, and $f$ is the plunge frequency. The reduced frequency $ k $ is defined as
	$k={2\pi f C }/{U_{\infty}}$, and dimensionless height $ h=Y/C $, where $ C $ is the chord length of the airfoil. In this study, $ k=2 $, $ h=0.4 $~\cite{Yu2016}. The whole mesh is considered as an oscillating rigid body. At the far field, $ \boldsymbol{q}^{p,b} = \boldsymbol{q}^{p,\infty} $, and the approximate Riemann solver is used to enforce boundary conditions.
	
\begin{figure}[]
	\centering
	\begin{tabular}{cc}
	\includegraphics[width=7cm]{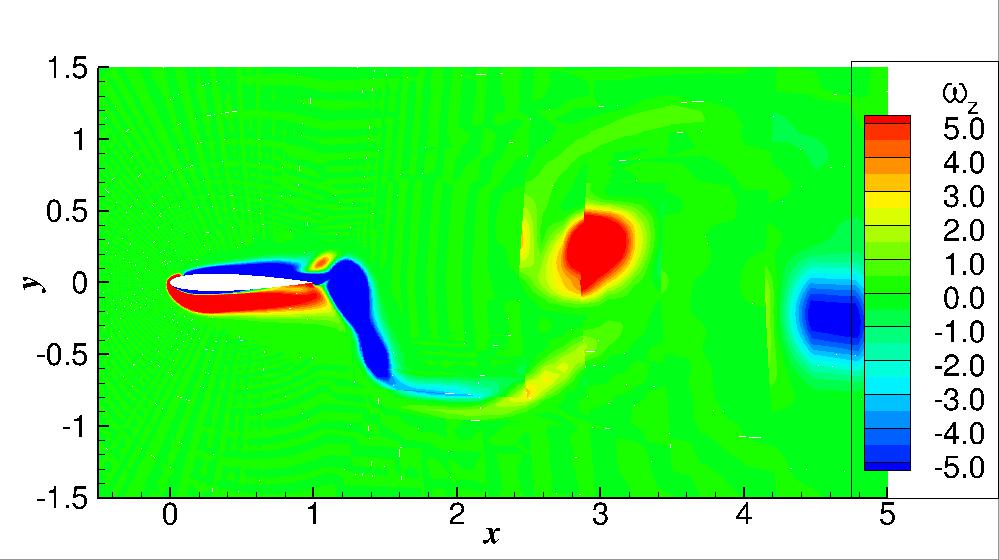}&
	\includegraphics[width=7cm]{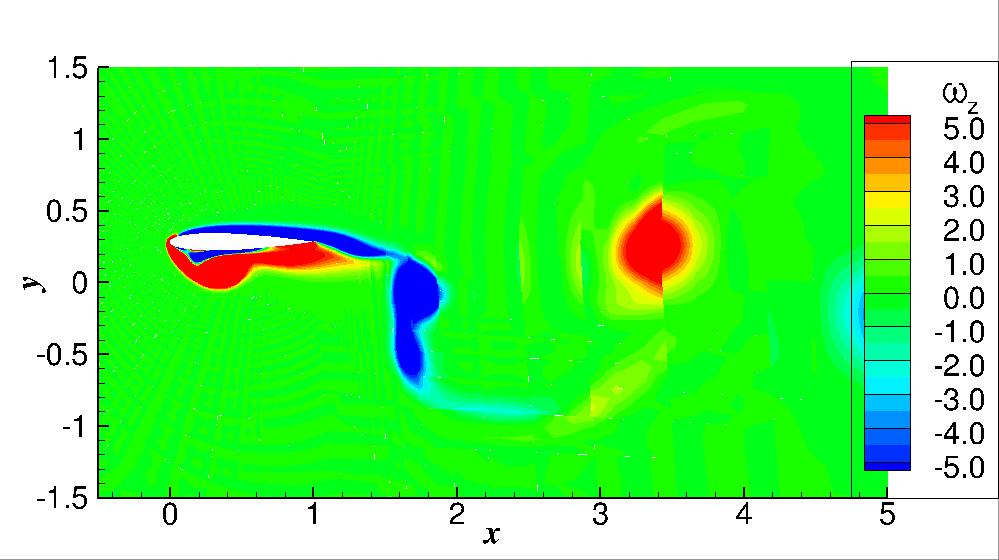}\\
	(a)&(b)\\
	\includegraphics[width=7cm]{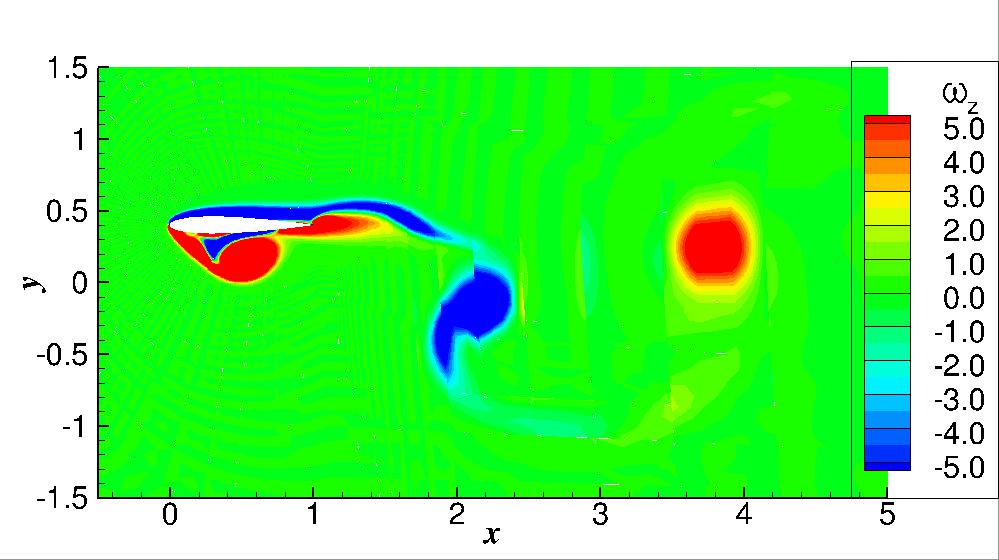}&
	\includegraphics[width=7cm]{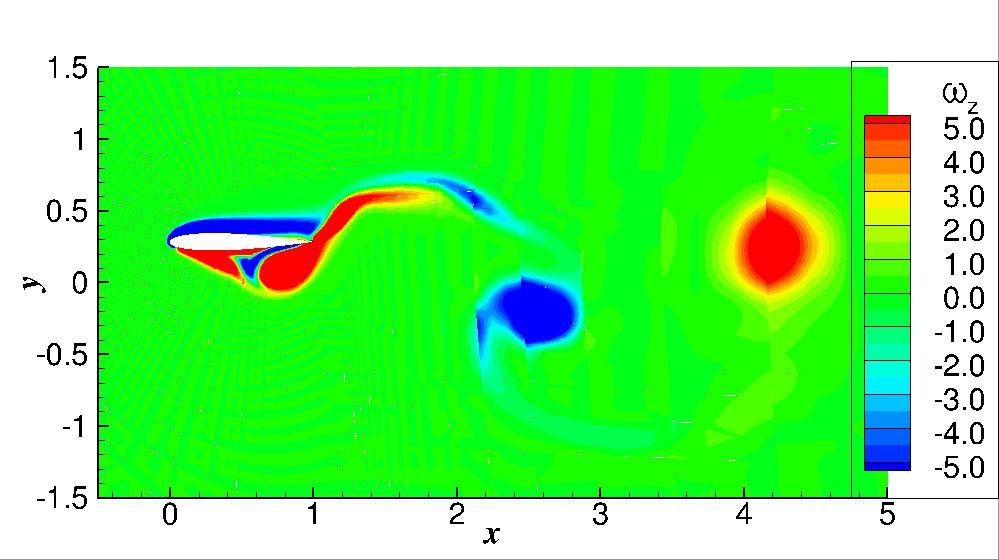}\\
	(c)&(d)\\
	\includegraphics[width=7cm]{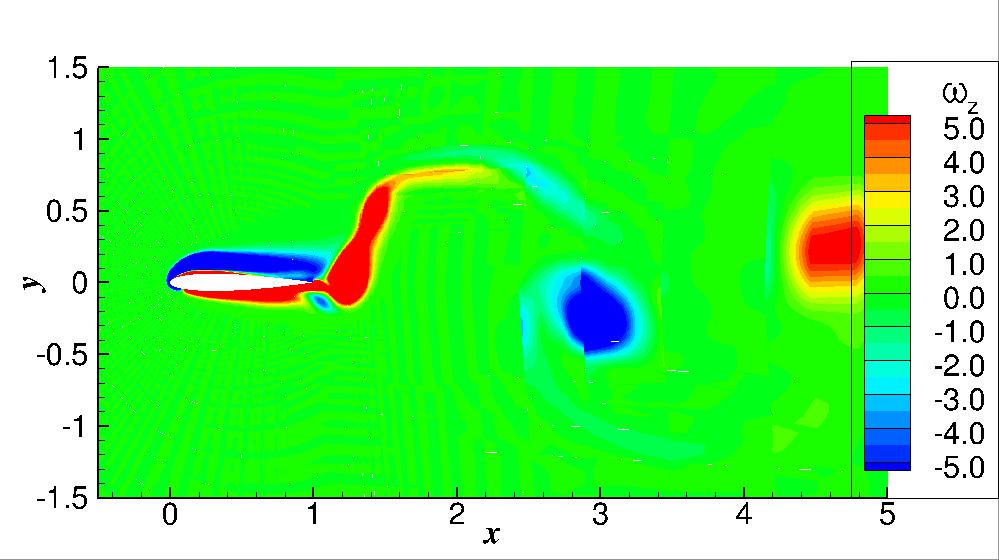}&
	\includegraphics[width=7cm]{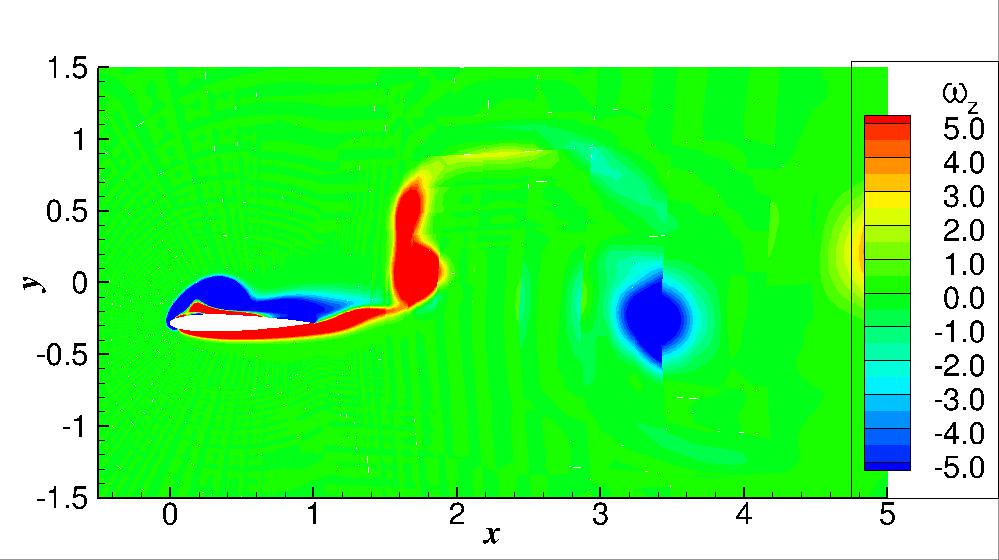}\\
	(e)&(f)\\
	\includegraphics[width=7cm]{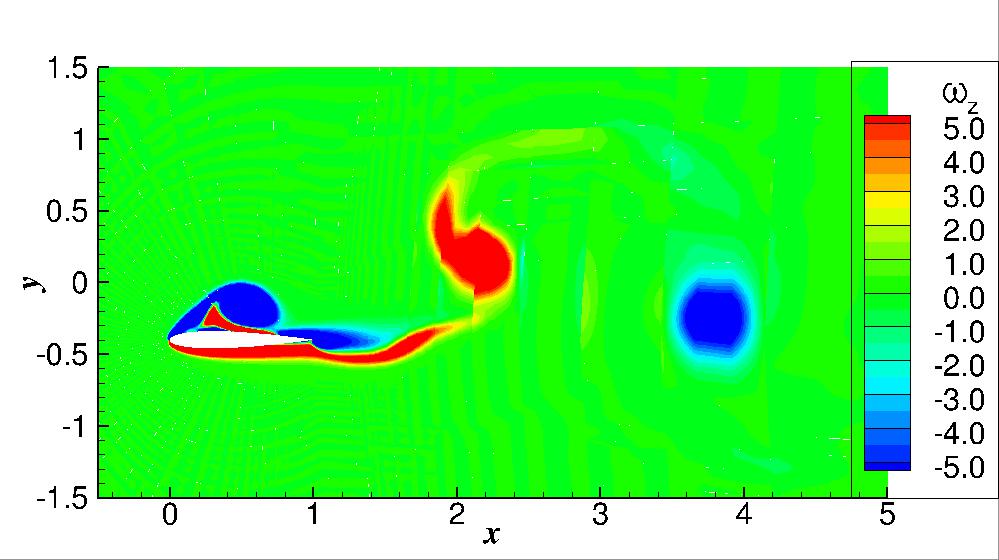}&
	\includegraphics[width=7cm]{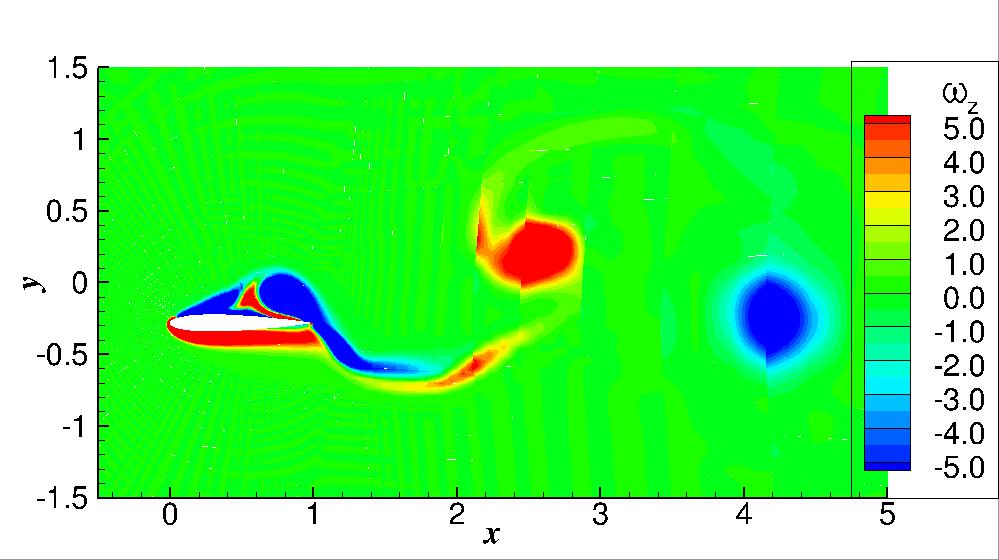}\\
	(g)& (h)\\
	\end{tabular}
	\caption{The vortex shedding process over a plunging NACA0012 airfoil in one oscillation cycle when the free stream Mach number $Ma_{\infty}$ is $ 10^{-3} $. (a) $0T $, (b) $ \frac{1}{8}T $, (c) $ \frac{2}{8}T $, (d) $ \frac{3}{8}T $, (e) $ \frac{4}{8}T $, (f) $ \frac{5}{8}T $, (g) $ \frac{6}{8}T $, and (h) $ \frac{7}{8}T $. The contour represents the vorticity in the direction (i.e., z-direction) perpendicular to the observation window.}
	\label{naca_vort_one_period}
	\end{figure}

\begin{figure}[]
	\centering
	\begin{tabular}{cc}
	\includegraphics[width=6cm]{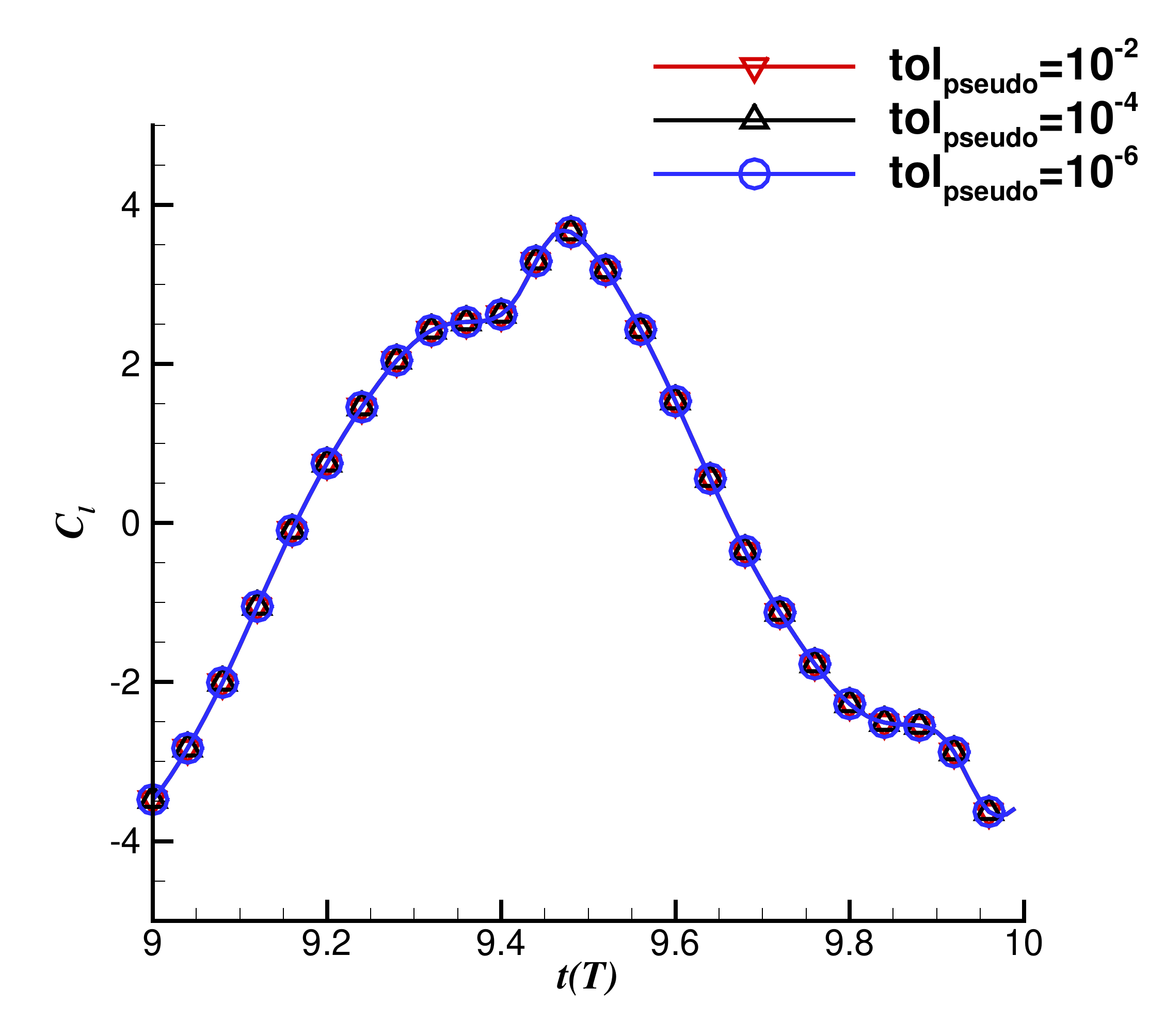} &
	\includegraphics[width=6cm]{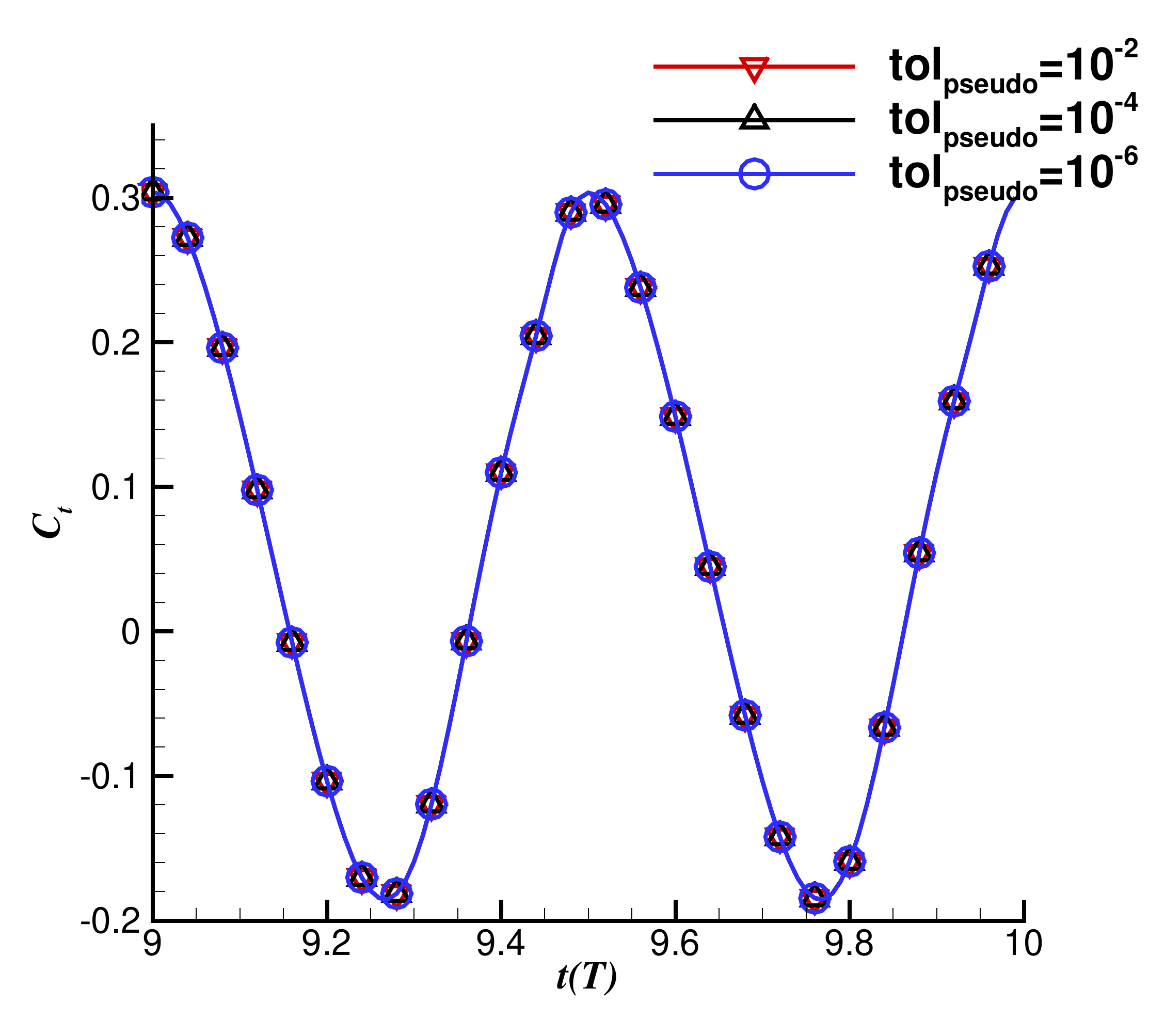}\\
	(a)& (b)\\
	\end{tabular}
	\caption{(a) $ C_l $ and (b) $ C_t $ of the tenth plunging period when different convergence criteria of $ Res_m/Res_0 $ are used in the simulation of laminar flow over the plunging airfoil at $ Ma_\infty = 10^{-3} $.}
	\label{naca_diff_tol_forces}
	\end{figure}

	The $ tol_{res} $ of the restarted GMRES solver is $ 10^{-6} $ with the maximum iteration number 100. For the sake of efficiency, we would require the $ Res_{m}/Res_{0} $ drop by certain orders instead of as low as possible. In this case, three criteria for $ Res_{m}/Res_{0} $ of the pseudo-time marching have been tested, i.e., $ tol_{pseudo} = 10^{-2} $, $ 10^{-4} $, and $ 10^{-6} $, to examine if the force predictions are sensitive to the convergence criterion of $ Res_m/Res_0 $. The time step size $ \Delta t $ is set to $ T/100 $, and $\kappa$ is set to $\sqrt{5}$. All simulations end at $ t_{end} = 10T $.  In Figure~\ref{naca_vort_one_period}, the vortex shedding process during one typical plunging period is presented. The thrust coefficient $ C_t $ and lift coefficient $ C_l $ of the tenth period are displayed in Figure~\ref{naca_diff_tol_forces}, and the time-averaged thrust coefficient $\bar{C_t}$, root mean square of the lift coefficient $C_{l,rms}$, and maximum lift coefficient $C_{l,max}$ are documented in Table~\ref{naca_diff_tol_forces_results}. From Figure~\ref{naca_diff_tol_forces} and Table~\ref{naca_diff_tol_forces_results}, we observe that a moderate tolerance ($ 10^{-4} $) for $Res_{m}/Res_{0}$ is sufficient to maintain the accuracy of force prediction. However, a smaller $ Res_m/Res_0 $ criterion means more iterations. 

	\begin{table}
	\centering
	\caption{Study on the effects of the convergence criterion $ tol_{pseudo} $ on force prediction of a plunging NACA0012 airfoil, $ \Delta t = T/100 $, and $ Ma_{\infty}=10^{-3} $.} 
	\label{naca_diff_tol_forces_results}
	\begin{tabular}{llll}
	\hline
	\hline
	 $ tol_{pseudo} $  & $ \bar{C_t} $ &$ C_{l,rms} $&$ C_{l,max} $\\
	\hline
	$ 10^{-2} $&0.0589&2.352& 3.685      \\
	$ 10^{-4} $&0.0589&2.351& 3.684 \\
	$ 10^{-6} $&0.0589&2.351& 3.684\\	
	\hline  
	\hline
	\end{tabular}
	\end{table}
	
	\subsubsection*{Effects of the Mach Number on Force Generation}
\begin{figure}[]
	\centering
	\begin{tabular}{cc}
	\includegraphics[width=6cm]{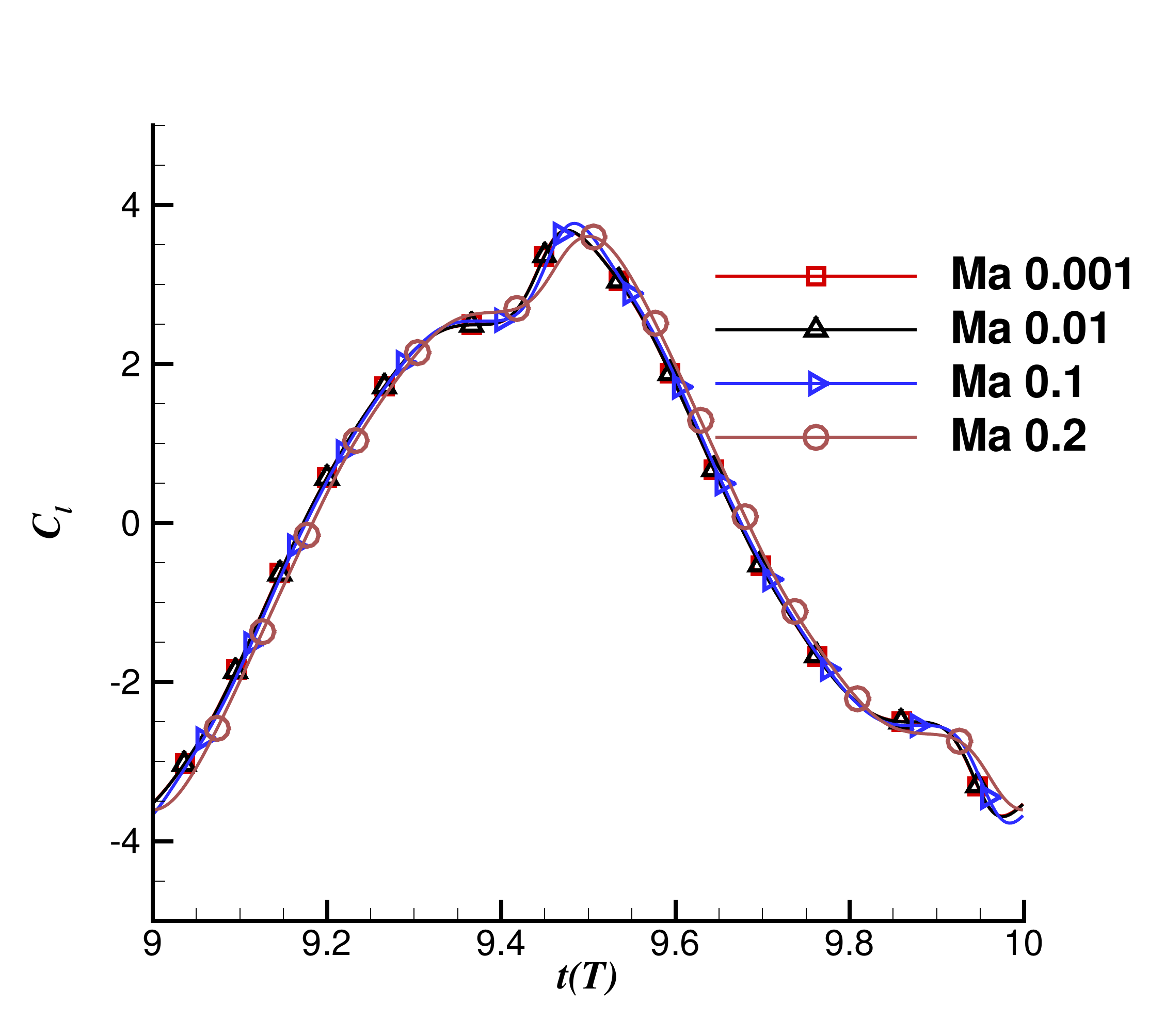} &
	\includegraphics[width=6cm]{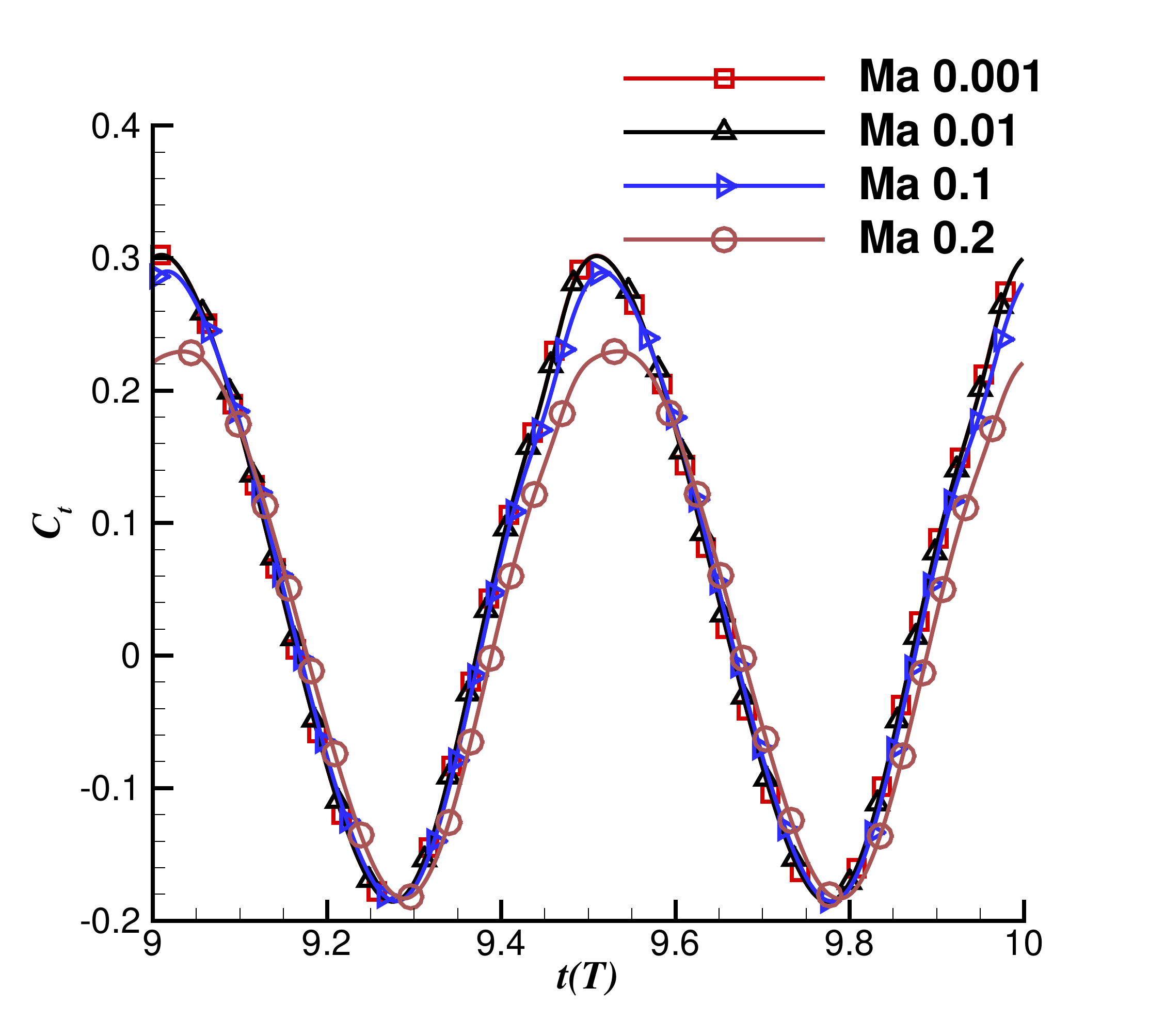}\\
	(a)& (b)\\
	\end{tabular}
	\caption{(a) $ C_l $ and (b) $ C_t $ of the tenth plunging period at different Mach numbers.}
	\label{diff_ma_forces}
	\end{figure}

	We further conducted a numerical study to test how the Mach number affects force prediction. Different Mach numbers, i.e., $ 10^{-3} $, $ 10^{-2} $, $ 10^{-1} $ and $ 0.2 $, are studied here.~A smaller time step size is employed, i.e.,~$ \Delta t=T/1000 $, to ensure that the time integration errors are negligible. $ tol_{pseudo} = 10^{-4} $ is adopted as the convergence criterion for pseudo-time iterations. The $ C_l $ and $ C_t $  of the tenth period are illustrated in Figure~\ref{diff_ma_forces}. $ \bar{C_{t}} $, $ C_{l,rms} $ and $ C_{l,max} $ are documented in Table~\ref{naca_diff_ma_forces_results}.
	\begin{table}
		\centering
		\caption{Study on the effects of the Mach number on force prediction of a plunging NACA0012 airfoil, $ \Delta t=T/1000 $.} 
		\label{naca_diff_ma_forces_results}
		\begin{tabular}{llll}
			\hline
			\hline
			$ Ma $  & $ \bar{C_t} $ &$ C_{l,rms} $&$ C_{l,max} $\\
			\hline
			$ 10^{-3} $&0.0578&2.340& 3.685      \\
			$ 10^{-2} $&0.0578&2.341& 3.692 \\
			$ 10^{-1} $&0.0539&2.361& 3.774\\	
			$ 0.2 $    &0.0370&2.369& 3.608\\
			Incompressible~\cite{Yu2016} &0.0557&2.348& \quad -\\
			\hline  
			\hline
		\end{tabular}
	\end{table}
	
	It is observed that when $ Ma $ is less than $ 10^{-2} $, the compressibility is almost negligible. The force histories for $ Ma=10^{-3} $ and $ Ma=10^{-2} $ coincide with each other. When $ Ma $ is larger than $ 10^{-1} $, phase shift is observed in the force histories. In general, when $ Ma $ is increased to a certain level (e.g., $ 10^{-1 }$ in this case), the time-averaged thrust coefficient $ \bar{C_t} $ will decrease when $ Ma $ increases; at the same time, the root mean square of the lift coefficient $ C_{l,rms} $ and the maximum lift coefficient $ C_{l,max} $ will vary slightly. For example, when $ Ma=0.2 $, the reduction of $ \bar{C_t} $ is over $ 36\% $ compared to that at $ Ma=10^{-3} $;  $ C_{l,rms} $ will increase about $ 1\% $; $ C_{l,max} $ will first increase and then decrease within a small range.
	 
	\section{Conclusions} \label{sec:conclusions}
	In this study, a high-order preconditioned FR/CPR method is developed to solve compressible Navier--Stokes equations at low Mach numbers on moving grids. The dual-time stepping method is used to handle unsteady flows. Specifically, BDF2 is adopted to discretize the physical time derivative term, and the restarted GMRES linear solver with the modified SER pseudo-time step size adaptation strategy is used to converge unsteady residuals at each physical time. For simulations with moving/deforming grids, the GCL is enforced implicitly to eliminate the discrepancy between the temporal and spatial discretization. The preconditioned FR/CPR method has demonstrated good accuracy and convergence through various benchmark tests. 
	
	We first study the convergence properties of the preconditioned FR/CPR method for steady inviscid flow simulations. Two cases, namely, low-Mach-number (i.e., $Ma = 10^{-3}$ in this study) flows over a cylinder and a NACA0012 airfoil are simulated. We observe that when the SER algorithm is used, the convergence criterion $ tol_{res} $ of the restarted GMRES solver should be sufficiently tight (or small) for the sake of convergence. When $ \Delta \tau \to \infty $ and SER is aggressive, the number of iterations for the GMRES solver to converge can be significantly increased, especially for higher-order spatial schemes on fine meshes. A remedy could be decreasing the maximum pseudo-time step $ \Delta \tau_{max} $. As a trade-off, the convergence speed of pseudo transient continuation will reduce correspondingly. A robust and efficient strategy for pseudo transient continuation is worthy of more research efforts in the future.
	
	The preconditioned FR/CPR method is then used to simulate low-Mach-number unsteady flows. The temporal and spatial orders of accuracy are first verified using isentropic vortex propagation on both static and dynamic meshes. After that, we study the numerical properties of the preconditioned FR/CPR method with low-Mach-number flows over a plunging NACA0012 airfoil. 
	We find that the convergence tolerance $tol_{pseudo}$ for $ Res_m/Res_0 $ of the pseudo-time marching does not need to be very small to preserve simulation accuracy. Since less pseudo-time iterations are needed when a larger criterion for $ Res_m/Res_0 $ is used, computational cost of unsteady flows can be decreased. We also conduct numerical experiments to test the effects of the Mach number on unsteady force generation. We find that when the Mach number is small enough (e.g., $ Ma = 10^{-2} $), flow compressibility has negligible effect on force prediction of a plunging airfoil. However, when the Mach number becomes larger (e.g., $ Ma = 0.2 $), thrust generation of a plunging airfoil will change significantly while the change of lift and its root mean square are not apparent.
	
	\section*{Acknowledgment}
	The authors gratefully acknowledge the support of the Office of Naval Research through
	the award N00014-16-1-2735, and the faculty startup support from the department of
	mechanical engineering at University of Maryland, Baltimore County (UMBC).
	\clearpage
	
	\bibliographystyle{ieeetr}
	\bibliography{LMP}

\begin{thebibliography}{10}

\bibitem{patankar1980numerical}
S.~Patankar, {\em Numerical Heat Transfer and Fluid Flow}.
\newblock CRC press, 1980.

\bibitem{chorin1968numerical}
A.~J. Chorin, ``{Numerical solution of the Navier-Stokes equations},'' {\em
  Mathematics of Computation}, vol.~22, no.~104, pp.~745--762, 1968.

\bibitem{chorin1967numerical}
A.~J. Chorin, ``{A numerical method for solving incompressible viscous flow
  problems},'' {\em Journal of Computational Physics}, vol.~2, no.~1,
  pp.~12--26, 1967.

\bibitem{Helenbrook06}
B.~T. Helenbrook, ``{Artificial compressibility preconditioning for
  incompressible flows under all conditions},'' in {\em the 44th Aerospace
  Sciences Meeting and Exhibit Conference}, (Reno, NV), 2006.
\newblock AIAA-2006-0689.

\bibitem{bassi2006artificial}
F.~Bassi, A.~Crivellini, D.~A. Di~Pietro, and S.~Rebay, ``{An artificial
  compressibility flux for the discontinuous Galerkin solution of the
  incompressible Navier--Stokes equations},'' {\em Journal of Computational
  Physics}, vol.~218, no.~2, pp.~794--815, 2006.

\bibitem{crivellini2013high}
A.~Crivellini, V.~D’Alessandro, and F.~Bassi, ``{High-order discontinuous
  Galerkin solutions of three-dimensional incompressible RANS equations},''
  {\em Computers \& Fluids}, vol.~81, pp.~122--133, 2013.

\bibitem{cox2016high}
C.~Cox, C.~Liang, and M.~W. Plesniak, ``{A high-order solver for unsteady
  incompressible Navier--Stokes equations using the flux reconstruction method
  on unstructured grids with implicit dual time stepping},'' {\em Journal of
  Computational Physics}, vol.~314, pp.~414--435, 2016.

\bibitem{Yu2016}
M.~L. Yu and L.~Wang, ``{A high-order flux reconstruction/correction procedure
  via reconstruction formulation for unsteady incompressible flow on
  unstructured moving grids},'' {\em {Computers \& Fluids}}, vol.~139,
  pp.~161--173, 2016.

\bibitem{TurkelLeer1993}
E.~Turkel, A.~Fiterman, and B.~Van~Leer, ``Preconditioning and the limit to the
  incompressible flow equations,'' tech. rep., Institute for Computer
  Applications in Science and Engineering, Hampton VA, 1993.

\bibitem{Colin2011}
{Y. Colin, H. Deniau and J-F. Boussuge}, ``{A robust low speed preconditioning
  formulation for viscous flow computations},'' {\em {Computers \& Fluids}},
  vol.~47, no.~1, pp.~1--15, 2011.

\bibitem{Turkel1987}
E.~Turkel, ``Preconditioned methods for solving the incompressible and low
  speed compressible equations,'' {\em Journal of Computational Physics},
  vol.~72, no.~2, pp.~277--298, 1987.

\bibitem{Turkel1997}
{E. Turkel, R. Radespiel and N. Kroll}, ``{Assessment of preconditioning
  methods for multidimensional aerodynamics},'' {\em {Computers \& Fluids}},
  vol.~26, no.~6, pp.~613--634, 1997.

\bibitem{Turkel_1999_review}
E.~Turkel, ``{Preconditioning techniques in computational fluid dymamics},''
  {\em Annual Review of Fluid Mechanics}, vol.~31, no.~1, pp.~385--416, 1999.

\bibitem{ChoiMerkle1993}
{Y. H. Choi and C. L. Merkle}, ``{The application of preconditioning in viscous
  flows},'' {\em Journal of Computational Physics}, vol.~105, pp.~207--233,
  1993.

\bibitem{WeissSmith1995}
J.~M. Weiss and W.~A. Smith, ``{Preconditioning applied to variable and
  constant density Flows},'' {\em AIAA Journal}, vol.~33, no.~11, 1995.

\bibitem{VanLeerRoe1991}
B.~van Leer, W.~Lee, and P.~Roe, ``{Characteristic time-stepping or local
  preconditioning of the Euler equations},'' in {\em the 10th Computational
  Fluid Dynamics Conference}, (Honolulu,HI), 1991.
\newblock AIAA-91-1552-CP.

\bibitem{LeeLeer1993}
D.~Lee and B.~van Leer, ``{Progress in local preconditioning of the Euler and
  Navier-Stokes equations},'' in {\em the 11th AIAA Computational Fluid
  Dynamics Conference}, (Orlando, FL), 1993.
\newblock AIAA-93-3328-CP.

\bibitem{Hauke1998}
G.~Hauke and T.~Hughes, ``{A comparative study of different sets of variables
  for solving compressible and incompressible flows},'' {\em Computer Methods
  in Applied Mechanics and Engineering}, vol.~153, pp.~1--44, 1998.

\bibitem{VenkaMerkle1995}
S.~Venkateswaran and L.~Merkle, ``{Analysis of preconditioning methods for the
  Euler and Navier-Stokes equations},'' in {\em {Lecture series-van Kareman
  Institute for fluid dynamics 3}}, 1995.
\newblock B1-B155.

\bibitem{Vigneron2006}
{D. Vigneron, G. Deli\'{e}ge and J. A. Essers}, ``{Low mach number local
  preconditioning for unsteady viscous finite volumes simulations on 3D
  unstructured grids},'' in {\em ECCOMAS CFD 2006: Proceedings of the European
  Conference on Computational Fluid Dynamics}, (Egmond aan Zee, The
  Netherlands), 2006.

\bibitem{Bassi2009}
{F. Bassi, C. De Bartolo, R. Hartmann and A. Nigro}, ``{A discontinuous
  Galerkin method for inviscid low Mach number flows},'' {\em Journal of
  Computational Physics}, vol.~228, pp.~3996--4011, 2009.

\bibitem{Luo2008}
{H. Luo, J. D. Baum and R. L\"{o}hner}, ``{A Fast $p$-Multigrid Discontinuous
  Galerkin Method for Compressible Flows at All Speeds},'' {\em AIAA Journal},
  vol.~46, no.~3, pp.~635--652, 2008.

\bibitem{Feistauer2007}
{M. Feistauer and V. Kucera}, ``{On a robust discontinuous Galerkin technique
  for the solution of compressible flow},'' {\em Journal of Computational
  Physics}, vol.~224, no.~1, pp.~208--221, 2007.

\bibitem{Nigro2010}
R.~H. A.~Nigro, C. De~Bartolo and F.~Bassi, ``{Discontinuous Galerkin solution
  of preconditioned Euler equations for very low Mach number flows},'' {\em
  International Journal for Numerical Methods in Fluids}, vol.~63,
  pp.~449--467, 2010.

\bibitem{Renda2015}
{S. M. Renda, R. Hartmann, C. De Bartolo and M. Wallraff}, ``{A high-order
  discontinuous Galerkin method for all-speed flows},'' {\em {International
  Journal for Numerical Methods in Fluids}}, vol.~77, pp.~224--247, 2015.

\bibitem{thornber2008improved}
B.~Thornber, A.~Mosedale, D.~Drikakis, D.~Youngs, and R.~J. Williams, ``{An
  improved reconstruction method for compressible flows with low Mach number
  features},'' {\em Journal of computational Physics}, vol.~227, no.~10,
  pp.~4873--4894, 2008.

\bibitem{rieper2011low}
F.~Rieper, ``{A low-Mach number fix for Roe’s approximate Riemann solver},''
  {\em Journal of Computational Physics}, vol.~230, no.~13, pp.~5263--5287,
  2011.

\bibitem{chen2018improved}
S.-S. Chen, C.~Yan, S.~Lou, and B.-X. Lin, ``{An improved entropy-consistent
  Euler flux in low Mach number},'' {\em Journal of Computational Science},
  vol.~27, pp.~271--283, 2018.

\bibitem{chen2018effective}
S.-S. Chen, C.~Yan, and X.-H. Xiang, ``Effective low-mach number improvement
  for upwind schemes,'' {\em Computers \& Mathematics with Applications},
  vol.~75, no.~10, pp.~3737--3755, 2018.

\bibitem{boxi2017study}
B.~Lin, C.~Yan, and S.~Chen, ``{A study on the behaviour of high-order flux
  reconstruction method with different low-dissipation numerical fluxes for
  large eddy simulation},'' {\em International Journal of Computational Fluid
  Dynamics}, vol.~31, no.~9, pp.~339--361, 2017.

\bibitem{Yu2011}
{M. L. Yu, Z. J. Wang and H. Hu}, ``{A high-order spectral difference method
  for unstructured dynamic grids},'' {\em {Computers \& Fluids}}, vol.~48,
  pp.~84--97, 2011.

\bibitem{Liang14}
C.~Liang, K.~Miyaji, and B.~Zhang, ``{An Efficient Correction Procedure via
  Reconstruction for Simulation of Viscous Flow on Moving and Deforming
  Domains},'' {\em Journal of Computational Physics}, vol.~256, pp.~55--68,
  2014.

\bibitem{Wang_Yu2017}
L.~Wang and M.~L. Yu, ``{A Preconditioned Flux Reconstruction/Correction
  Procedure via Reconstruction Formulation for Unsteady Low {Mach} Number Flows
  on Dynamic Unstructured Meshes},'' in {\em the 55th AIAA Aerospace Sciences
  Meeting}, (Grapevine, TX), 2017.
\newblock AIAA-2017-0737.

\bibitem{Pletcher1993}
R.~H. Pletcher, ``{On solving the compressible Navier-Stokes equations for
  unsteady flows at very low Mach numbers},'' in {\em the 11th Computational
  Fluid Dynamics Conference}, 1993.
\newblock AIAA-93-3368-CP.

\bibitem{Jameson1991}
A.~Jameson, ``{Time dependent calculations using multigrid, with applications
  to unsteady flows past airfoils and wings},'' in {\em the 10th Computational
  Fluid Dynamics Conference}, (Honolulu,HI), 1991.

\bibitem{Xiao2007}
{T. H. Xiao, H. S. Ang and S. Z. Yu}, ``{A preconditioned dual time-stepping
  procedure coupled with matrix-free LU-SGS scheme for unsteady low speed
  viscous flows with moving objects},'' {\em {International Journal of
  Computational Fluid Dynamics}}, vol.~21, no.~3-4, pp.~165--173, 2007.

\bibitem{Huynh2007}
{H. T. Huynh}, ``{A flux reconstruction approach to high-Order schemes
  including discontinuous Galerkin methods},'' in {\em the 18th AIAA
  Computational Fluid Dynamics Conference}, (Miami, FL), 2007.
\newblock AIAA-2007-4079.

\bibitem{Huynh2009}
{H. T. Huynh}, ``{A reconstruction approach to high-order schemes including
  discontinuous Galerkin methods for diffusion},'' in {\em the 47th AIAA
  Aerospace Sciences Meeting including The New Horizons Forum and Aerospace},
  (Orlando, FL), 2009.
\newblock AIAA-2009-403.

\bibitem{ZJWang2009}
{Z. J. Wang and H. Y. Gao}, ``{A unifying lifting collocation penalty
  formulation including the discontinuous Galerkin, spectral volume/difference
  methods for conservation laws on mixed grids},'' {\em Journal of
  Computational Physics}, vol.~98, pp.~209--220, 2014.

\bibitem{Vincent2011}
P.~E. Vincent, P.~Castonguay, and A.~Jameson, ``{A new class of high-order
  energy stable flux reconstruction schemes},'' {\em Journal of Scientific
  Computing}, no.~1, pp.~50--72, 2011.

\bibitem{BR2_2005}
{F. Bassi and S. Rebay}, ``Discontinuous {Galerkin} solution of the
  {Reynolds-averaged Navier-Stokes} and k-{$\omega$} turbulence model
  equations,'' {\em {Computers \& Fluids}}, vol.~34, pp.~507--540, 2005.

\bibitem{Gao2013}
{H. Y. Gao, Z. J. Wang and H. T. Huynh}, ``{Differential formulation of
  discontinuous Galerkin and related methods for the Navier-Stokes
  equations},'' {\em {Communications in Computational Physics}}, vol.~{13},
  no.~4, pp.~1013--1044, 2013.

\bibitem{Mulder1985}
{W. A. Mulder and B. van Leer}, ``{Experiments with Implicit Upwind Methods for
  the Euler Equations},'' {\em Journal of Computational Physics}, vol.~59,
  pp.~232--246, 1985.

\bibitem{ceze2011robust}
M.~Ceze and K.~Fidkowski, ``{A robust adaptive solution strategy for high-order
  implicit CFD solvers},'' in {\em 20th AIAA Computational Fluid Dynamics
  Conference}, p.~3696, 2011.

\bibitem{ceze2015constrained}
M.~Ceze and K.~J. Fidkowski, ``Constrained pseudo-transient continuation,''
  {\em International Journal for Numerical Methods in Engineering}, vol.~102,
  no.~11, pp.~1683--1703, 2015.

\bibitem{petsc-user-ref}
S.~Balay, S.~Abhyankar, M.~F. Adams, J.~Brown, P.~Brune, K.~Buschelman,
  L.~Dalcin, A.~Dener, V.~Eijkhout, W.~D. Gropp, D.~Kaushik, M.~G. Knepley,
  D.~A. May, L.~C. McInnes, R.~T. Mills, T.~Munson, K.~Rupp, P.~Sanan, B.~F.
  Smith, S.~Zampini, H.~Zhang, and H.~Zhang, ``{PETS}c users manual,'' Tech.
  Rep. ANL-95/11 - Revision 3.10, Argonne National Laboratory, 2018.

\bibitem{Bassi2015}
{F. Bassi, L. Botti, A. Colombo, A Ghidoni and F. Massa}, ``{Linearly implicit
  Rosenbrock-type Runge-Kutta schemes applied to the Discontinuous Galerkin
  solution of compressible and incompressible unsteady flows},'' {\em
  {Computers \& Fluids}}, vol.~118, no.~2, pp.~305--320, 2015.

\bibitem{Hesthaven08}
J.~S. Hesthaven and T.~Warburton, {\em Nodal discontinuous Galerkin methods –
  algorithms, analysis and applications}.
\newblock New York: Springer, 2008.

\end{thebibliography}
\end{document}